%% file: main.tex
\documentclass[aps, reprint, prb, twocolumn, amsmath, amssymb, groupedaddress, 10pt, longbibliography]{revtex4-1}

\usepackage{epsfig}
\usepackage{subfigure}
\usepackage{graphicx}
\usepackage{amsfonts}
\usepackage[figuresright]{rotating}
\usepackage{amssymb}
\usepackage{amsmath}
\usepackage{psfrag}
\usepackage{esint} 
\usepackage{hyperref}
\hypersetup{colorlinks,allcolors=blue} 
\usepackage{multirow}
\usepackage{siunitx}
\usepackage{xcolor}
\usepackage{lipsum}
\usepackage{bbm}
\usepackage{soul}

\global\hyphenpenalty=100000

\setcitestyle{super,open={},close={}}

\begin{document}

\title{Anisotropic Josephson coupling of \texorpdfstring{$d$}{d} vectors in triplet superconductors 
arising 
from frustrated spin textures
}

\author{Grayson R. Frazier}
\affiliation{Department of Physics and Astronomy, Johns Hopkins University, Baltimore, Maryland 21218, USA}
\author{Junyi Zhang}
\affiliation{Department of Physics and Astronomy, Johns Hopkins University, Baltimore, Maryland 21218, USA}
\author{Yi Li}
\affiliation{Department of Physics and Astronomy, Johns Hopkins University, Baltimore, Maryland 21218, USA}

\date{\today}

\begin{abstract}
    We demonstrate that coupling itinerant electrons to a noncollinear classical exchange field can induce anisotropic Josephson coupling between superconducting $d$ vectors,  analogous to the Dzyaloshinskii-Moriya and $\Gamma$-type interactions in magnetism. 
    Using perturbative methods, we analyze an $s$-$d$ model on a geometrically frustrated lattice.
    Noncollinear local spin textures generate spin triplet pairing correlations and can favor spatially varying superconducting order due to anisotropic Josephson couplings between $d$ vectors,
    endowing a ``pliability'' to the pairing order that competes with the superfluid stiffness.
    For nonunitary pairing, this spatial texture of $d$ vectors can give rise to anomalous vortices in the absence of an external magnetic field. 
    We further predict a Josephson diode effect with efficiency proportional to the spin chirality of the underlying magnetic texture. 
    These results establish a link between frustrated magnetism and spatial textures of triplet superconducting pairing, with implications for a range of materials such as Mn$_3$Ge and $4H_b$-TaS$_2$, where superconductivity can be proximity-induced or intrinsic.
\end{abstract}

\maketitle

\input{sections/01.intro}

\input{sections/02.free_energy}

\input{sections/03.model}
\input{sections/04.effective_tunneling}

\input{sections/05.josephson_effects}
\input{sections/06.conclusion}

\input{main.bbl}


\end{document}


\title{Supplemental Material for ``Anisotropic Josephson coupling of \texorpdfstring{$d$}{d} vectors in triplet superconductors 
arising 
from 
frustrated spin textures''}

\author{Grayson R. Frazier}
\affiliation{Department of Physics and Astronomy, Johns Hopkins University, Baltimore, Maryland 21218, USA}
\author{Junyi Zhang}
\affiliation{Department of Physics and Astronomy, Johns Hopkins University, Baltimore, Maryland 21218, USA}
\author{Yi Li}
\affiliation{Department of Physics and Astronomy, Johns Hopkins University, Baltimore, Maryland 21218, USA}


\maketitle

\onecolumngrid

\setcounter{equation}{0}
\setcounter{section}{0}
\setcounter{figure}{0}
\setcounter{table}{0}
\setcounter{page}{1}
\makeatletter
\renewcommand{\theequation}{S\arabic{equation}}
\renewcommand{\thefigure}{S\arabic{figure}}
\renewcommand{\thesection}{S\arabic{section}}

\input{supplemental/s1.josephson_tunneling}
\input{supplemental/s2.model}
\input{supplemental/s3.pairing_correlations}
\input{supplemental/s4.effective_tunneling}

\input{supplemental/s5.effective_spin_exchange}

\input{supplemental/s6.vortices}
\input{supplemental/s7.majorana}
\input{supplemental/s8.large_Jsd}

\input{supplemental.bbl}

%% file: sections/01.intro.tex

Recent experimental results have suggested interesting interplay between unconventional superconductivity and frustrated magnetic textures.
For example, in the triangular antiferromagnetic (AFM) Weyl semimetal 
Mn$_3$Ge~\cite{Kiyohara2016, Liu2017, Wuttke2019, Dasgupta2020},
correlating 
with the presence of an anomalous Hall effect in its chiral noncollinear AFM phases, long-range coherent Josephson supercurrents 
have been observed, 
implying proximity-generated spin triplet pairing~\cite{Jeon2021}.
Furthermore, an out-of-plane canted magnetization, corresponding to finite spin chirality of local spin moments in Mn$_3$Ge, has been shown to produce hysteretic Josephson supercurrents~\cite{Jeon2023}.
More recently, experiments on $4H_b$-TaS$_2$, consisting of alternating centrosymmetric $1T$-TaS$_2$ spin liquid candidate layer~\cite{Fazekas1979, Law2017, Ribak2017, Klanjsek2017, He2018, ManasValero2021, Vano2021} 
and noncentrosymmetric $1H$-TaS$_2$ superconducting layer~\cite{Nagata1992},
have demonstrated 
chiral superconducting states, spontaneous vortices, and the ``magnetic memory" effect in addition to spin triplet pairing~\cite{Ribak2020, Persky2022, Silber2024}.
In both of these materials, frustrated magnetic textures are widely believed to underlie these phenomena
\cite{Lin2024, Liu2024, Koenig2024, Crippa2024, Levitan2025, Frazier2025b},
though the interplay between the frustrated spin textures and unconventional superconductivity requires further study.

The interplay between frustrated magnetism and unconventional superconductivity has been studied via a range of theoretical approaches, from microscopic Bogoliubov-de~Gennes (BdG) equations~\cite{Bogoliubov1958, deGennes1966} to quasiclassical Andreev equations~\cite{Andreev1964a} for wavefunction based methods, and from Gor'kov equations~\cite{Gorkov1958, Nambu1960} of the Green's functions of a microscopic many-body Hamiltonian to {\color{black}Eilenberger-Larkin-Ovchinnikov and Usadel equations}~\cite{Eilenberger1968, Larkin1969, Usadel1970}.
While BdG formalism captures the microscopic pairing mechanisms, it is limited by computational cost for complex real-space spin textures. 
Quasiclassical methods efficiently handle proximity-induced pairing correlations but typically assume weak gradients and smooth spin textures. 
In this work, we first discuss a general form of the free energy contribution from a spatially varying $d$ vector order parameter describing spin triplet pairing correlations.
We find that an anisotropic coupling between $d$ vectors is allowed when the effective Josephson tunneling occurs in the presence of, for example, a local exchange field composed of frustrated spins.
We next develop a complementary microscopic framework 
using Green's function methods based on $T$-matrix formalism
that conveniently incorporates frustrated spin textures into effective Josephson couplings between $d$ vectors via a controlled perturbative expansion.
To demonstrate, we consider a simple $s$-$d$ model on a three-sublattice system, in which the itinerant electrons are coupled to a local exchange field.
The local exchange field leads to an effective tunneling which is dependent on the underlying frustration of the spin structure.
Ultimately, this {\color{black} results in} an anisotropic Josephson coupling of spin triplet pairing orders between adjacent grains, leading to a ``pliability'' of the order parameter that competes with the superfluid stiffness and promotes a spatially inhomogeneous $d$ vector texture.
Lastly, we discuss the appearance of anomalous vortices in addition to a spin-chirality induced Josephson diode effect.


%% file: sections/02.free_energy.tex
\paragraph*{Contribution to free energy from a spatially varying order parameter.---} 
Generally, the free energy can be separated into contributions from homogeneous pairing orders and gradient terms associated with the cost of spatial variations of the pairing order.
%
%
%
For a conventional $s$-wave superconductor, 
the latter contribution is given by $F_\mathrm{variation} = \gamma \int \mathrm{d}^d r \, |\boldsymbol{\nabla} \psi(\mathbf{r})|^2$, with $\psi(\mathbf{r})$ being the pairing order and $\gamma$ the superfluid stiffness.
The discretized form of $F_\mathrm{variation}$ describes the Josephson coupling of weakly-linked grains, $\sum_{\langle n m\rangle} J_{nm} \psi_n \psi^*_m + \mathrm{c.c.},$
where $\psi_{n}$ is the uniform local pairing order of the $n\mathrm{th}$ grain.
$J_{n m}$, the lattice-version of the superfluid stiffness, is the Josephson coupling amplitude
between adjacent grains $n$ and $m$~\cite{Spivak1991, Sigrist1995}.
Typically, $J_{nm}$ is real and negative, which favors a phase-coherent, homogeneous pairing order in sufficiently large grains where Coulomb charging energy is negligible.

Now, let us consider spin triplet superconductors where the spin degree of freedom of the pairing order is encoded in its $d$ vector~\cite{Leggett1975, Volovik2009, Alicea2012, Cornfeld2021}, which is a complex vector with its real and imaginary part normal to the spin of the Cooper pair. 
The Josephson coupling between two spin triplet superconductors {\color{black}depends} not only on the phase difference but also on the orientations of $d$ vectors.
We first examine the simplest case of a generalized $^3$He-A-type pairing, 
where the spin part is characterized by a $d$ vector decoupled 
from the orbital degree of freedom in the pairing order.
The free energy contribution from the Josephson coupling is given by
\begin{eqnarray}
        F_{n m}
        &=&
        e^{i(\phi_n - \phi_m)}
        \Big\{ J_{n m} \hat{\mathbf{d}}_{n} \cdot \hat{\mathbf{d}}_{m}^*
        +
        \mathbf{D}_{n m} \cdot \Big( \hat{\mathbf{d}}_{n} \times \hat{\mathbf{d}}_{m}^* \Big)
        \nonumber \\
        &&+
        \sum\nolimits_{a,b = 1,2,3}
        \hat{d}_{n}^{a} \Gamma_{n m; ab} \hat{d}^{*b}_{m}
        \Big\} 
        + \mathrm{c.c.}
    \label{josephson_coupling.d_vectors}
\end{eqnarray}
Here, $\hat{\mathbf{d}}_{n} = (\hat{d}_{n}^{1},\hat{d}_{n}^{2},\hat{d}_{n}^{3})$ denotes the $d$ vector  of the pairing order at $n\mathrm{th}$ grain, with each component being a complex function, and 
$\phi_n$ is the $\mathrm{U}(1)$ superconducting phase at that grain. 
%
The first term $J_{n m}$ corresponds to a Heisenberg-like symmetric coupling between $d$ vectors and favors
collinear alignment of $d$ vectors between adjacent grains. 
The second $\mathbf{D}_{n m}$ term, analogous to Dzyaloshinskii-Moriya (DM) interaction~\cite{Dzyaloshinsky1958, Moriya1960, Moriya1960a, Hellman2017, Hill2021} in magnetism, 
is antisymmetric and describes anisotropic DM-like coupling of $d$ vectors. 
The third term $\Gamma_{n m}$ is a symmetric traceless rank-$2$ tensor, analogous to ``$\Gamma$-type'' interaction~\cite{Rau2014} in magnetism.
The competition of these three terms 
{\color{black}can lead to a frustrated $d$ vector texture}.
%
In the continuum limit, the free energy in Eq.~\eqref{josephson_coupling.d_vectors} is given by
\begin{math}
    F_\mathrm{variation} = \int \mathrm{d}^dr [ \sum_i \gamma (\mathbf{r}) \partial_i \hat{\mathbf{d}}(\mathbf{r}) \cdot \partial_i \hat{\mathbf{d}}^*(\mathbf{r})
    +
    \sum_{i} \mathbf{A}_i (\mathbf{r}) \cdot (\partial_i \hat{\mathbf{d}}(\mathbf{r}) \times \hat{\mathbf{d}}^*(\mathbf{r})) 
    + 
    \sum_{ij} K_{ij} (\mathbf{r}) \partial_i \hat{\mathbf{d}} (\mathbf{r}) \cdot  \partial_j \hat{\mathbf{d}}^*(\mathbf{r})
    ]
    +
    \mathrm{c.c.}.
\end{math}
The last two terms give rise to an effective ``pliability" of the pairing order, competing with the superfluid stiffness in the first term and potentially driving a reorientation of $d$ vectors towards a spatially varying texture.

The first order Josephson free energy in Eq.~\eqref{josephson_coupling.d_vectors} can be derived from the Ambegaokar-Baratoff formalism microscopically. It takes the form $F_{n m} = 
\mathrm{Re} (\mathcal{J}_{n m})$ 
where the Josephson form factor $\mathcal{J}_{nm}$ is given by~\cite{Ambegaokar1963, Geshkenbein1986, Sigrist1991, Frazier2024}
\begin{equation}
    \mathcal{J}_{nm}=
    - \frac{1}{\beta} \sum_{i\omega_n}
    \mathrm{Tr} \Big[ \mathcal{F}_{n}(i\omega_n) [T_{n m}(i\omega_n)]^\mathrm{T}
    \mathcal{F}_{m}^*(i\omega_n)
    T_{n m}(i\omega_n) \Big].
    \label{form_factor}
\end{equation}
Here, $\mathcal{F}_{\mu \nu} (\mathbf{k}; i\omega_n) 
= - \int \mathrm{d} \tau \,  e^{i\omega_n \tau}
\langle \mathcal{T}_\tau c_{-\mathbf{k}, \nu} (\tau) c_{\mathbf{k}, \mu}(0) \rangle$
is the Matsubara frequency representation of the pairing correlation in the $n\mathrm{th}$ superconducting grain. 
When there is only spin triplet pairing, the anomalous Green's function can be characterized by the $d$ vector,
$\mathcal{F}_{n} \sim (\hat{\mathbf{d}}_n \cdot \boldsymbol{\sigma}) (i\sigma^y)$, 
where $\boldsymbol{\sigma} = (\sigma^x, \sigma^y, \sigma^z)$ are the Pauli matrices. 
$T_{n m}=\sum_{i\in n, j\in m}T_{ij}$, where $T_{ij}$ is the tunneling matrix that describes the effective tunneling of single electrons between the $i\mathrm{th}$ and $j\mathrm{th}$ sites of the $n\mathrm{th}$ and  $m\mathrm{th}$ grains, respectively~\cite{Bardeen1961, Cohen1962, Caroli1971, Caroli1971a}.
The trace is taken over internal degrees of freedom, {\color{black}such as spin and orbital, }
and the summation is over fermion Matsubara frequencies, $\omega_n = (2n+1)\pi/\beta$. 
The total Josephson energy describing the spatial variation of the pairing order is given by the summation of individual contributions between adjacent grains, $ \sum_{\langle n m \rangle} F_{n m}$.
For systems with broken inversion symmetry,
the Josephson form factor in Eq.~\eqref{form_factor} can admit couplings between spin singlet and triplet pairing correlations in presence of spin-dependent tunneling~\cite{SM}.

The quadratic couplings of $d$ vectors in the Josephson free energy in Eq.~\eqref{josephson_coupling.d_vectors} are proportional to tunneling amplitudes $T_{nm;0}$ and $\mathbf{T}_{n m}=(T_{nm;1},T_{nm;2},T_{nm;3})^\mathrm{T}$ as~\cite{Millis1988, Sigrist1991, Frazier2024}
\begin{align}
        &J_{ n m} \propto 
        T_{n m; 0}^2 + \mathbf{T}_{n m} \cdot \mathbf{T}_{n m},
        \hspace{1em}
        \mathbf{D}_{n m} \propto 
        i T_{n m; 0} \mathbf{T}_{n m}, \nonumber
        \\
        &\Gamma_{n m;ab} \propto 
        T_{n m; a} T_{n m; b}.
    \label{Josephson_couplings_Tmn}
\end{align}
Here, the tunneling matrix is decomposed into its spin-independent and spin-dependent tunneling processes,
\begin{math}
    T_{n m}  = T_{n m; 0} \sigma^0 + \mathbf{T}_{n m} \cdot \boldsymbol{\sigma}.
\end{math}
The spin-dependent effective tunneling can arise in the presence of spin-orbit coupling, or when time-reversal symmetry is broken, due to, for example, coupling with local exchange fields.
In the former case, 
the Heisenberg-like, DM-like, and $\Gamma$-type Josephson couplings in Eq.~\eqref{josephson_coupling.d_vectors} are generally momentum-dependent and describe the Josephson couplings for $d$ vector textures over Fermi surfaces in grains $n$ and $m$~\cite{Frazier2025b, SM}.

%% file: sections/03.model.tex
\paragraph*{Effective tunneling in presence of a local exchange field~--}
To obtain the anisotropic Josephson couplings, we consider a minimal $s$-$d$ model on a geometrically frustrated system, such as a two-dimensional kagome or triangular lattice.
Itinerant $s$ electrons are coupled to the local spins of $d$ electrons,  
which play the role of a local exchange field.
The 
Hamiltonian is given by
\begin{math}
    H
    = H_{\mathrm{kin}} + H_{sd},
\end{math}
in which the kinetic term $H_{\mathrm{kin}}$ describes a system with spin-independent nearest-neighbour hopping,
\begin{math}
    H_{\mathrm{kin}} = t_0 \sum_{\langle \mathbf{r}_i, \mathbf{r}_j \rangle, \alpha}
    c^\dagger_{\mathbf{r}_i, \alpha}
    c_{\mathbf{r}_j, \alpha}
    -\mu \sum_{\mathbf{r}_i, \alpha} c^\dagger_{\mathbf{r}_i, \alpha} c_{\mathbf{r}_i, \alpha}.
    \label{kinetic.NN.real_space}
\end{math}
Here, $t_0<0$ is the hopping amplitude, $\mu$ the chemical potential, and $c^\dagger_{\mathbf{r}_i, \alpha}$ the creation operator for $s$ electrons at site $\mathbf{r}_i$ and spin $\alpha = {\uparrow, \downarrow}$.
At the mean-field level, the $s$-$d$ exchange is described by
\begin{math}
    H_{sd} = 
    J_{sd}
    \sum_{\mathbf{r}_j, \alpha, \alpha'}
    c^{\dagger}_{\mathbf{r}_j, \alpha}
    [
    \boldsymbol{\sigma}_{\alpha, \alpha'}
    \cdot
    \mathbf{s}_j
    ]
    c_{\mathbf{r}_j, \alpha'},
    \label{sd_coupling_ham.real_space}
\end{math}
in which
$J_{sd}$ is the amplitude of the $s$-$d$ coupling~\cite{Anderson1961, Kondo1962, Kondo1964, Schrieffer1966, Chen2014, Coleman2015}.
The local exchange field is described by the spin moment ${\mathbf{s}}_j = \langle d^\dagger_{\mathbf{r}_j, \alpha} \boldsymbol{\sigma}_{\alpha, \alpha'} d_{\mathbf{r}_j, \alpha'} \rangle/2$, with $d^\dagger_{\mathbf{r}_j, \alpha}$ being the creation operator for the $d$ electrons.
For the remainder of this work, we treat these local spins as classical fields.

\begin{figure} 
    \centering
    \includegraphics[width=\linewidth]{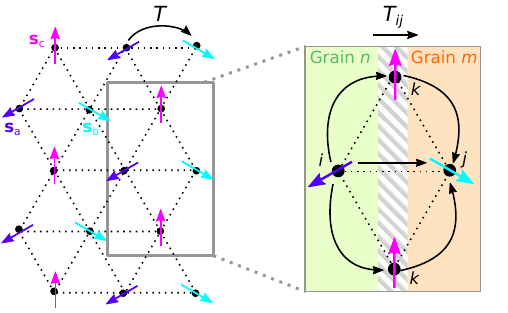}
    \caption{Schematic of the effective tunneling and Josephson coupling.
    (Left) Triangular lattice with local spin moments $\mathbf{s}_a$, $\mathbf{s}_b$ and $\mathbf{s}_c$ (purple, cyan, pink) which form the three-sublattice unit cell.
    Between two sites, there is an effective tunneling $T_{ij}$ of itinerant electrons due to the local spin moments.
    (Right)
    Zoomed in view of the tunneling between sites $i$ and $j$.
    One can treat the tunneling between adjacent sites $i$ and $j$ as a microscopic Josephson junction at the interface between grains $n$ and $m$ described by pairing order parameters $\mathbf{d}_{n}$ and $\mathbf{d}_{m}$ respectively.
    In addition to the spin-independent tunneling, there is a higher order tunneling process that is mediated by neighbouring site labelled by $k$. 
    A similar picture applies to the kagome lattice.
    }
    \label{fig:tunneling}
\end{figure}

%% file: sections/04.effective_tunneling.tex
{
\color{black}
To see explicitly the role of the local exchange field in the tunneling of itinerant electrons, we perform a $T$-matrix-like expansion~\cite{Shiba1968, Rusinov1969, Balatsky2006} of the $s$-$d$ coupling.
}
The effective tunneling between neighbouring sites $i$ and $j$ in the presence of a local spin texture is given by~\cite{SM}
\begin{align}
    T_{i j}(i\omega)
    &=
    \ t_0 
    \sigma^0
    +
    J_{sd}^2
    [\mathcal{G}_\mathrm{kin}(i\omega)]_{i j}
    \alpha_{ij}
    \sigma^0 \nonumber
    \\
    &+
    i J_{sd}^2
    [\mathcal{G}_\mathrm{kin}(i\omega)]_{i j}
    \boldsymbol{\beta}_{ij} \cdot
    \boldsymbol{\sigma} \nonumber
    \\
    &- 
    i J_{sd}^3
    \sum_{k}
    [\mathcal{G}_\mathrm{kin}(i\omega)]_{i k}
    [\mathcal{G}_\mathrm{kin}(i\omega)]_{k j}
    \chi_{ijk}
    \sigma^0 \nonumber
    \\
    &+
    J_{sd}^3
    \sum_{k}
    [\mathcal{G}_\mathrm{kin}(i\omega)]_{i k}
    [\mathcal{G}_\mathrm{kin}(i\omega)]_{k j}
    \boldsymbol{\gamma}_{ijk}
    \cdot
    \boldsymbol{\sigma}. 
\label{effective_tunneling_Jsd_expansion}
\end{align}
Above, 
the summation is taken over intermediate sites $k$ that are nearest neighbours to sites $i$ and $j$.
We have defined the following factors that depend on the local exchange field,
\begin{equation}
    \begin{gathered}
        \alpha_{ij} \equiv \mathbf{s}_{i} \cdot \mathbf{s}_{j};
        \hspace{2em}
        \boldsymbol{\beta}_{ij} \equiv \mathbf{s}_{i} \times \mathbf{s}_{j};
        \\
        \chi_{ijk} \equiv \mathbf{s}_{i} \cdot (\mathbf{s}_{j}\times \mathbf{s}_{k});
        \\
        \boldsymbol{\gamma}_{ijk} \equiv
        (\mathbf{s}_{i} \cdot \mathbf{s}_{k})\mathbf{s}_{j}
        -
        (\mathbf{s}_{i} \cdot \mathbf{s}_{j})\mathbf{s}_{k}
        +
        (\mathbf{s}_{j} \cdot \mathbf{s}_{k})\mathbf{s}_{i}.
    \end{gathered}
    \label{spin_factors}
\end{equation}
For a three sublattice system, $\mathbf{s}_{i}$, $\mathbf{s}_{j}$, and the intermediate site's spin $\mathbf{s}_{k}$ correspond to the three spins $\mathbf{s}_{a}$, $\mathbf{s}_b$, and $\mathbf{s}_c$, as depicted in Fig.~\ref{fig:tunneling}.
Above, $\alpha_{ij}$ is invariant under inversion and time reversal symmetries, whereas $\boldsymbol{\beta}_{ij}$ is invariant under time reversal but odd under inversion along the tunneling direction.
In contrast, the third-order term $\boldsymbol{\gamma}_{ijk}$ is invariant under inversion yet breaks time reversal symmetry.
Only the spin chirality term, $\chi_{ijk}$, 
{\color{black}breaks both inversion and time reversal} symmetries~\cite{Wen1989}.
The terms $\boldsymbol{\beta}_{ij}$, $\boldsymbol{\gamma}_{ijk}$, and $\chi_{ijk}$ can be nonvanishing for a {\color{black}noncollinear}, frustrated spin texture.
%
{
\color{black}
In this work, we are primarily interested in the role of Josephson coupling of spin triplet Cooper pairs, even for coplanar spin structures with vanishing scalar spin chirality.
Hence, we do not include the minimal coupling to the gauge field arising from nonvanishing spin chirality.
}


%% file: sections/05.josephson_effects.tex
\paragraph*{Josephson exchange couplings.---}
We now examine the free energy terms in Eq.~\eqref{josephson_coupling.d_vectors} using the effective tunneling arising from the coupling to frustrated local spin moments.
Specifically, we consider two adjacent grains of the superconductor, shown schematically in Fig.~\ref{fig:tunneling}~(Right), in which the tunneling between two grains predominantly occurs at the grain boundary.
The contribution to the Josephson free energy between two adjacent grains is related to the form factor in Eq.~\eqref{form_factor}.  
Here, the effective tunneling matrices $T_{n m}$ are given by Eq.~\eqref{effective_tunneling_Jsd_expansion}, which incorporates the effects of the local exchange field.
Because tunneling is determined by the sites at the grain boundary, the tunneling process will depend on the local spin configuration at the interface.
In the following, we make the approximation that the Josephson coupling arises primarily from effective nearest neighbour tunneling between boundary sites of adjacent grains.
The total Josephson coupling is obtained by summing over all such tunneling processes along the grain boundary.

From the effective tunneling in Eq.~\eqref{effective_tunneling_Jsd_expansion}, the Josephson couplings in Eq.~\eqref{Josephson_couplings_Tmn}, to third order in $J_{sd}$, are approximated by~\cite{SM}
\begin{eqnarray}
    J_{n m} 
    &\approx& \nonumber
    \Delta_0
    \sum_{i \in \Sigma_n,j \in \Sigma_m}\Big(
    - \frac{t_0^2}{W^2} 
    {+} \frac{J_{sd}^2}{ W^2} \alpha_{i j}  
    + 4i \frac{J_{sd}^3}{W^2 t_0} \chi_{i j k} 
    \Big),
    \\
    \mathbf{D}_{n m} 
    &\approx &
    \Delta_0
    \sum_{i \in \Sigma_n,j \in \Sigma_m}\Big(
    {-2} \frac{J_{sd}^2}{W^2} \boldsymbol{\beta}_{i j} 
    - 4i \frac{J_{sd}^3}{W^2t_0} \boldsymbol{\gamma}_{i j k} 
    \Big).
    \label{Jnm_Dnm_three_sublattice}
\end{eqnarray}
Here, $W$ denotes the band width, and 
the summation of $i$ and $j$ is taken over the grain boundaries $\Sigma_n$ and $\Sigma_m$ of the $n\mathrm{th}$ and $m\mathrm{th}$ grains respectively.
We have already summed over nearest neighbouring sites $k$, which denotes the third sublattice mediating the higher order tunneling process between sublattices $i$ and $j$, as shown schematically in Fig.~\ref{fig:tunneling}~(Right).
The leading order term of the symmetric tensor $\Gamma_{n m}$ in Eq.~\eqref{Josephson_couplings_Tmn} is fourth order in $J_{sd}$ and is {\color{black} omitted here}~\cite{SM}.
Above, we have used the approximation that $\mathcal{G}_\mathrm{kin} \approx-1/t_0$ for nearest neighbours, and that the BdG quasiparticle energy is set by the pairing amplitude, $\Delta_0$.
This assumption is valid for the bands at the Fermi level, which primarily contribute to the Josephson tunneling.

\begin{figure}
    \centering
    \includegraphics[width=\linewidth]{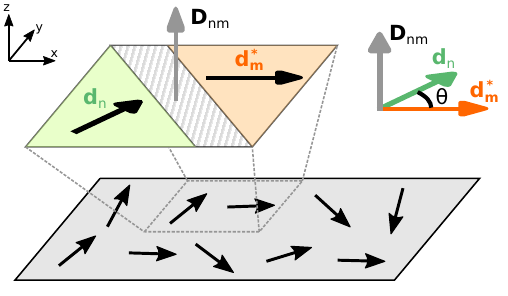}
    \caption{
    Zoomed-in view of example $d$ vector texture (black) arising from anisotropic Josephson exchange couplings.
    Noncollinear texture of the underlying spin moments at the grain boundary (gray, dashed) give rise to a DM-like Josephson coupling $\mathbf{D}_{n m}$ and Heisenberg-like coupling $J_{nm}$ between grains with order parameters described by $d$ vectors $\mathbf{d}_{n}$ and $\mathbf{d}_{m}$.
    The relative angle between $\mathbf{d}_{n}$ and $\mathbf{d}_{m}^*$ is given by the relative magnitude of $J_{n m}$ and $\mathbf{D}_{n m}$ in Eq.~\eqref{Jnm_Dnm_three_sublattice}.
    }
    \label{fig:d-vector_texture}
\end{figure}

In the limit of vanishing $J_{sd},$ only the Heisenberg-like coupling $J_{n m}$ is nonvanishing, leading to collinear $d$ vectors at adjacent grains and thereby promoting a spatially uniform $d$ vector texture.
A similar outcome arises for nonfrustrated magnetic textures (\textit{i.e.} collinear spin configurations), in which the stiffness term $J_{n m}$ dominates.
However, for frustrated noncollinear magnetic textures with finite $J_{sd}$, both the Heisenberg-like and DM-like Josephson couplings in Eq.~\eqref{josephson_coupling.d_vectors} compete with one another, favoring collinear and noncollinear alignment of $d$ vectors, respectively.
The relative angle $\theta$ of $d$ vectors at adjacent grains is given by $\tan \theta = D_\perp/J_{n m}$, in which $D_\perp = \mathbf{D}_{n m} \cdot (\hat{\mathbf{d}}_{n} \times \hat{\mathbf{d}}_{m}^*)/|\hat{\mathbf{d}}_{n} \times \hat{\mathbf{d}}_{m}^*|$ is the component of $\mathbf{D}_{n m}$ normal to the plane spanned by $\mathbf{d}_{n}$ and $\mathbf{d}_{m}^*$.
The misalignment of $d$ vectors, as illustrated in Fig.~\ref{fig:d-vector_texture}, 
{\color{black} can give} rise to topological textures such as vortices, spirals, or skyrmions in the $d$ vector field.
The nonvanishing anisotropic Josephson coupling can be interpreted as endowing a ``pliability'' to the pairing order, which competes with the stiffness imposed by the Heisenberg-like coupling.
Hence, in the presence of a frustrated exchange field, the system minimizes its Josephson free energy by developing a spatially inhomogeneous $d$ vector texture.
It is important to note that a nonvanishing scalar spin chirality is not required to realize the anisotropic DM-like Josephson coupling.
For instance, a $120^\circ$ ordered coplanar spin structure already breaks the necessary symmetries to realize the anisotropic Josephson coupling.

{\color{black}
The relative magnitudes of the Heisenberg-like and DM-like couplings can give insight to characteristic length scales of the $d$ vector textures.
For representative parameters relevant to $4H_b$-TaS$_2$~\cite{Vano2021}, we take $W \sim 8t_0$ and $J_{sd} \sim t_0/10$, and we consider the $120^\circ$ ordered antiferromagnetic spin configuration, in which $\boldsymbol{\beta}_{ij} = (\sqrt{3}/2)\epsilon_{ij}\hat{z}$, with $\epsilon_{ij}$ being the antisymmetric tensor.
From Eq.~\eqref{Jnm_Dnm_three_sublattice}, the ratio of the two Josephson couplings is $|\mathbf{D}_{nm}|/|J_{nm}| \approx 0.02$,
leading to a relative tilt of $\theta \approx 0.02$ between $d$ vectors at adjacent grains.
The characteristic length of $d$ vector textures is given by $\lambda = (2\pi/\theta)\xi$, in which $\xi$ is the size of superconducting grains, with lower bound set by the superconducting coherence length.
For $\xi$ on the order of tens of nanometers~\cite{Ribak2020, Jeon2021}, this yields a lower bound of $\lambda$
on the order of tens of microns, corresponding to, for example, $d$ vector vortex textures with diameter on the order of a few microns.
For systems with stronger $s$-$d$ coupling---such as  Mn$_3$Ge, in which the $s$-$d$ coupling can be comparable to the hopping amplitude~\cite{Chen2014, Kimata2019,Khadka2020}---the effective tunneling in Eq.~\eqref{effective_tunneling_Jsd_expansion} can admit higher order terms; nonetheless, the Josephson couplings in Eq.~\eqref{josephson_coupling.d_vectors} persist~\cite{Frazier2025b}.
}

\paragraph*{Anomalous vortices and spin chirality induced Josephson diode effect.---}
The spatially inhomogeneous $d$ vector texture arising from the anisotropic Josephson coupling can lead to anomalous vortices, even in the absence of a magnetic field or when spin chirality of the underlying local exchange field is zero.
For spin triplet pairing order described by 
\begin{math}
    \Delta_j (\mathbf{r}) = |\Delta_j(\mathbf{r})| \hat{\Delta}_j(\mathbf{r})
\end{math}
in which
\begin{math}
    \hat{\Delta}_j = e^{i\phi(\mathbf{r})} \hat{d}_j (\mathbf{r}) \sigma^j (i\sigma^y),
\end{math}
the superfluid velocity is given by
\begin{math}
    \mathbf{v}_s = - \frac{i}{2} \frac{\hbar}{m^*}
    ( \hat{\Delta}^\dagger_{j}(\mathbf{r}) \boldsymbol{\nabla} \hat{\Delta}_{j}(\mathbf{r}) - \mathrm{h.c.}),
\end{math}
with the summation implied over repeated index $j=1,2,3$ and $m^*$ being the mass of the Cooper pair.
{\color{black}Here, we disregard any sublattice or orbital degrees of freedom and only focus on the spin degree of freedom encoded in the $d$ vector for simplicity.}
The superfluid velocity includes a contribution from the spatial variation of the $\mathrm{U}(1)$ phase and also from the spatial variation of the $d$ vector,
\begin{math}
    \color{black}
    v_{s,j} = ({\hbar}/{m^*})
    (
    \partial_j \phi(\mathbf{r})
    -i \hat{\mathbf{d}}^*(\mathbf{r}) \cdot \partial_j \hat{\mathbf{d}}(\mathbf{r}) 
    ).
\end{math}
The latter contribution plays the role of a gauge connection and is nonvanishing for nonunitary pairing orders, which satisfy $\hat{\mathbf{d}}(\mathbf{r}) \times \hat{\mathbf{d}}^*(\mathbf{r}) \neq 0$.
Thus, coupling to a frustrated local exchange field promotes spatially inhomogeneous $d$ vector textures that lead to nontrivial contributions to the superfluid velocity.

Analogous to Mermin-Ho vortices~\cite{Mermin1976}, which arise from nontrivial circulation of the $l$ vector in superfluid $^3$He-A, nontrivial circulation of the $d$ vector can support coreless vortices.
The nontrivial circulation of velocity, in the absence of a magnetic field, is given by~\cite{Salomaa1987, Volovik2009, SM}
\begin{align}
    \color{black}
    (\boldsymbol{\nabla} \times \mathbf{v}_s)_i
    =
    \frac{1}{2}
    \frac{\hbar}{m^*}
    \epsilon_{ijk}
    \epsilon_{abc}
    \hat{S}_a \partial_j \hat{S}_b \partial_k \hat{S}_c,
    \label{curl_of_superfluid_velocity}
\end{align}
{\color{black}in which} $\hat{\mathbf{S}} = i\hat{\mathbf{d}}\times \hat{\mathbf{d}}^*/|\hat{\mathbf{d}}\times \hat{\mathbf{d}}^*|$ is the real unit vector pointing in the local direction of the {\color{black}Cooper pair's spin}.
{\color{black}
Even in the absence of an external magnetic field or singularities in the $\mathrm{U}(1)$ phase,
}
the DM-like coupling {\color{black} in Eq.~\eqref{josephson_coupling.d_vectors}} can promote noncollinear nonunitary $d$ vectors across neighbouring grains, giving rise to, for example, skyrmion-like defects of the $d$ vector which manifest as {\color{black}anomalous} vortices.

{ \color{black}
When there is additionally nontrivial circulation in the $\mathrm{U}(1)$ superconducting phase, unitary pairing order can support half-quantum vortices~\cite{Volovik2000, Ivanov2001, Babaev2002, Volovik2009}, in which the half-winding of the real $d$ vector is accompanied by a $\pi$-phase winding in the overall $\mathrm{U}(1)$ phase.
For nonunitary pairing order, in which $\hat{\mathbf{d}}$ is intrinsically complex, a rotation $\hat{\mathbf{d}} \rightarrow \hat{\mathbf{d}}' = e^{i\theta} \hat{\mathbf{d}}$
is equivalent to the pairing order acquiring $\mathrm{U}(1)$ phase winding, $\theta$.
}

Lastly, we propose that a superconductor-insulator-superconductor (SIS) Josephson junction, in which the barrier region has nonvanishing spin chirality, will display a Josephson diode effect proportional to the spin chirality.
{
\color{black}
The first order Josephson critical current at zero voltage bias between grains $n$ and $m$ is given by
\begin{math}
    I_J^{(n, m)}
    =
    i \frac{e}{\hbar} \left( \mathcal{J}_{nm} - \mathrm{c.c.}\right).
\end{math}
}
The directional dependent contribution to the Josephson supercurrent is given by~\cite{Frazier2025b}
\begin{equation}
    {
    \color{black}
    \delta I_c
    \propto
    \frac{\Delta_0 J_{sd}^3}{W^2 t_0} 
    \underset{j \in \Sigma_m}{\sum_{i \in \Sigma_n,}}
    \bigg\{
    \chi_{ijk}
    (\hat{\mathbf{d}}_n \cdot \hat{\mathbf{d}}_m)
    -
    \boldsymbol{\gamma}_{ijk} \cdot 
    (\hat{\mathbf{d}}_n \times \hat{\mathbf{d}}_m)
    \bigg \},
    }
    \label{nonreciprocal_current}
\end{equation}
{\color{black}in which 
$\delta I_c$ is the difference in the maximal and minimal critical currents.}
{\color{black}Intuitively, the Josephson diode effect arises from breaking parity and time reversal symmetries~\cite{Zhang2022a, Wang2025}.
The first contribution scales linearly with the spin chirality $\chi_{ijk}$, which breaks time-reversal and local inversion symmetry at the junction interface, and similarly, the second contribution is dependent on the time-reversal breaking term $\gamma_{ijk}$ in Eq.~\eqref{spin_factors} and the parity-breaking term $\hat{\mathbf{d}}_n \times \hat{\mathbf{d}}_m$.
}


Additionally, the system can host Majorana states localized at the junction interface which can lead to, for example, $4\pi$-periodic Josephson current~\cite{Kitaev2001, Kwon2003, Fu2008, Chung2009, Sau2010, Alicea2010, Grosfeld2011, Alicea2012, SM}.
The additional contribution to the critical current can affect the diode efficiency, but nonetheless, a finite diode effect can persist, provided that time-reversal and parity symmetries are broken at the junction interface.
Lastly, in the regime in which $J_{sd} \gtrsim t_0$, this can modify the Josephson coupling amplitudes in Eq.~\eqref{Jnm_Dnm_three_sublattice}, which can alter the Josephson diode efficiency~\cite{Frazier2025b, SM}.



%% file: sections/06.conclusion.tex
\paragraph*{Conclusion and Discussion.---}
We studied how a frustrated spin texture couples with itinerant electrons that form an unconventional spin triplet superconducting pairing order.
Analogous to Dzyaloshinskii-Moriya's anisotropic superexchange, anisotropic Josephson couplings between $d$ vectors of adjacent superconducting grains can contribute to the free energy,
which lead to
a ``pliability'' of the pairing order that opposes the superfluid stiffness, so that a spatially varying $d$ vector texture is energetically favored.
The inhomogeneous $d$ vector texture can lead to novel phenomena, {\color{black}including} coreless $d$ vector vortices and a Josephson diode effect.


Our results are pertinent to recent experimental demonstration of the coexistence of frustrated magnetic textures and superconductivity.
For example, in Mn$_3$Ge, noncollinear spin textures act as an intrinsic exchange field, and {\color{black}it} has been shown to go superconducting in proximity to Nb.
The noncollinear texture can lead to DM-like Josephson couplings that favor inhomogeneous $d$ vector textures.
Furthermore, Mn$_3$Ge 
can be used in a Josephson junction in the $ab$-plane to realize the Josephson diode effect.
By applying a magnetic field along the $c$-axis, the $120^\circ$ ordered spin configuration is canted out of plane, generating nonvanishing spin chirality and therefore a finite Josephson diode effect.

Another example is superconducting $4H_b$-TaS$_2$.
Noncollinear spin textures in the $1T$-TaS$_2$ layers can lead to anisotropic Josephson couplings.
The resultant $d$ vector textures can lead to spontaneous vortices at zero field.
The dynamics of the local spin moments 
in the $1T$-TaS$_2$ layers 
and their effect on the superconducting pair correlations remain a future avenue to be explored.

\vspace{0.5em}
\noindent
\paragraph*{Acknowledgments~--}
We are grateful for helpful suggestions from Oleg Tchernyshyov.
We acknowledge the support of the NSF CAREER Grant No.~DMR-1848349. 
JYZ acknowledges partial support from the Johns Hopkins University Theoretical Interdisciplinary Physics and Astronomy Center.

%% file: main.bbl
%

%% file: supplemental/s1.josephson_tunneling.tex
\section{Josephson form factor and free energy}
\label{SM:josephson}

\subsection{Josephson form factor}

We briefly review the Josephson form factor, which describes the Josephson coupling between two superconducting grains.
A more detailed derivation based on the Ambegaokar-Baratoff formalism of the Josephson current~\cite{Ambegaokar1963, Sigrist1991} can be found in, for example, Ref.~\citenum{Frazier2024}.

Consider two superconducting grains connected by a weak link.
The tunneling across the barrier between grains $n$ and $m$ can be expressed as
\begin{equation}
    [H_{T}]_{n m} = 
    \sum_{\mathbf{k}, \mathbf{k}'; \alpha, \alpha'} \left(  c^\dagger_{n, \mathbf{k}, \alpha} [T_{n m}(\mathbf{k}, \mathbf{k}')]_{\alpha \alpha'} c_{m, \mathbf{k}', \alpha'}
    + \mathrm{h.c.} \right),
\end{equation}
in which $c_{n, \mathbf{k}, \alpha}$ is the annihilation operator for an electron at grain $n$ with momentum $\mathbf{k}$ and spin $\alpha$.
The tunneling matrix element between grains $n$ and $m$ generally takes the form
\begin{math}
    T_{n m}(\mathbf{k}, \mathbf{k}')
    = T_{n m; 0}(\mathbf{k}, \mathbf{k}')  \sigma^0 + \mathbf{T}_{n m}(\mathbf{k}, \mathbf{k}')  \cdot \boldsymbol{\sigma},
\end{math}
in which $\boldsymbol{\sigma} = (\sigma^x, \sigma^y, \sigma^z)^\mathrm{T}$ are the Pauli matrices and $T_{n m; 0}(\mathbf{k}, \mathbf{k}') $ and $\mathbf{T}_{n m}(\mathbf{k}, \mathbf{k}') $ correspond to spin-independent and spin-dependent tunneling processes at the grain boundary.
The latter can arise from, for example, spin-orbit coupling or time-reversal breaking fields at the interface.

The Josephson form factor describes the tunneling of Cooper pairs and corresponds to the anomalous part of the current-current correlation function.
At zero voltage bias, the form factor is given by
\begin{equation}
    \mathcal{J}_{n m}=
    - \frac{1}{\beta}
    \sum_{i\omega_n}
    \sum_{\mathbf{k}, \mathbf{k}'}
    \mathrm{Tr}
    \Big[ \mathcal{F}_{n}(\mathbf{k}; i\omega_n) [T_{n m}(-\mathbf{k}, -\mathbf{k}'; i\omega_n)]^\mathrm{T}
    [\mathcal{F}_{m}^\dagger(\mathbf{k}; i\omega_n')]^\mathrm{T} T_{n m}(\mathbf{k}, \mathbf{k}'; i\omega_n)
    \Big],
    \label{appendixEq:form_factor}
\end{equation}
in which the trace is taken over internal degrees of freedom of the Cooper pair (\textit{e.g.} spin, sublattice, \textit{etc.}), and the summation is taken over fermionic Matsubara frequencies.
Here, $\mathcal{F}_n(\mathbf{k}; i\omega_n)$ is the anomalous Green's function of grain $n$.
Focusing on the spin degrees of freedom, the anomalous Green's function can be expanded into its spin singlet and spin triplet components as
\begin{equation}
    \mathcal{F}_{n}(\mathbf{k}; i\omega_n)
    =
    \Big( f_{n, 0} (\mathbf{k}; i\omega_n) + \mathbf{f}_n(\mathbf{k}; i\omega_n) \cdot \boldsymbol{\sigma} \Big)
    i\sigma^y.
\end{equation}
Substituting into Eq.~\eqref{appendixEq:form_factor}, the trace can be expressed as three distinct contributions,
\begin{equation}
    \mathrm{Tr}\Big[ \cdots \Big]
    =
    2
    \Big(
    C_{\mathrm{sing}-\mathrm{sing}}
    +
    C_{\mathrm{sing}-\mathrm{trip}}
    +
    C_{\mathrm{trip}-\mathrm{trip}}
    \Big),
\end{equation}
which correspond to the Josephson tunneling between spin singlet components, between spin triplet components, and between spin singlet and spin triplet components respectively.
The three contributions to the Josephson form factor are given by 
\begin{subequations}
    \begin{equation}
    \begin{aligned}
        C_{\mathrm{sing}-\mathrm{sing}}
        =
        &
        - f_{n, 0}(\mathbf{k}')
        f^*_{m, 0}(\mathbf{k}) 
        T_{n m; 0}(-\mathbf{k}, -\mathbf{k}')
        T_{n m; 0}(\mathbf{k}, \mathbf{k}')
        + f_{n, 0}(\mathbf{k}') f^*_{m, 0}(\mathbf{k})  \Big[ \mathbf{T}_{n m}(-\mathbf{k}, -\mathbf{k}') \cdot \mathbf{T}_{n m}(\mathbf{k}, \mathbf{k}')\Big]
    \end{aligned}
    \end{equation}
    \begin{equation}
    \begin{aligned}
        C_{\mathrm{sing}-\mathrm{trip}}
        =
        & \
        \Big[ \mathbf{f}_n(\mathbf{k}') \cdot \mathbf{T}_{n m}(-\mathbf{k}, -\mathbf{k}')\Big] f^*_{m, 0}(\mathbf{k}) T_{n m; 0}(\mathbf{k}, \mathbf{k}')
        +
        f_{n, 0}(\mathbf{k}') T_{n m; 0}(-\mathbf{k}, -\mathbf{k}') \Big[ \mathbf{f}_m^*(\mathbf{k}) \cdot \mathbf{T}_{n m}(\mathbf{k}, \mathbf{k}') \Big]
        \\
        &
        - T_{n m; 0}(-\mathbf{k}, -\mathbf{k}') f^*_{m, 0}(\mathbf{k}) \Big[ \mathbf{f}_n(\mathbf{k}') \cdot \mathbf{T}_{n m}(\mathbf{k}, \mathbf{k}')\Big]
        - 
        f_{n, 0}(\mathbf{k}') T_{n m; 0}(\mathbf{k}, \mathbf{k}')
        \Big[\mathbf{f}_m^*(\mathbf{k}) \cdot \mathbf{T}_{n m}(-\mathbf{k}, -\mathbf{k}')\Big]
        \\
        &+if_{n, 0}(\mathbf{k}') \mathbf{f}_m^*(\mathbf{k}) \cdot \Big[ \mathbf{T}_{n m}(-\mathbf{k}, -\mathbf{k}') \times \mathbf{T}_{n m}(\mathbf{k}, \mathbf{k}')\Big]
        + if^*_{m, 0}(\mathbf{k}) \mathbf{f}_n(\mathbf{k}') \cdot \Big[\mathbf{T}_{n m}(-\mathbf{k}, -\mathbf{k}') \times \mathbf{T}_{n m}(\mathbf{k}, \mathbf{k}')\Big]
    \end{aligned}
    \end{equation}
    \begin{equation}
    \begin{aligned}
        C_{\mathrm{trip}-\mathrm{trip}}
        = 
        & \
        T_{n m; 0}(-\mathbf{k}, -\mathbf{k}')  T_{n m; 0}(\mathbf{k}, \mathbf{k}') \Big[ \mathbf{f}_n(\mathbf{k}') \cdot \mathbf{f}_m^*(\mathbf{k})\Big]
        + 
        \Big[ \mathbf{T}_{n m}(-\mathbf{k}, -\mathbf{k}') \cdot \mathbf{T}_{n m}(\mathbf{k}, \mathbf{k}') \Big]
        \Big[ \mathbf{f}_n(\mathbf{k}') \cdot\mathbf{f}_m^*(\mathbf{k}) \Big]
        \\
        &
        + iT_{n m; 0}(\mathbf{k}, \mathbf{k}') \mathbf{T}_{n m}(-\mathbf{k}, -\mathbf{k}') \cdot \Big[ \mathbf{f}_n(\mathbf{k}') \times \mathbf{f}_m^*(\mathbf{k}) \Big]
        +
        iT_{n m; 0}(-\mathbf{k}, -\mathbf{k}')  \mathbf{T}_{n m}(\mathbf{k}, \mathbf{k}') \cdot \Big[ \mathbf{f}_n(\mathbf{k}') \times \mathbf{f}_m^*(\mathbf{k})  \Big]
        \\
        &
        -
        \Big[\mathbf{f}_n(\mathbf{k}') \cdot \mathbf{T}_{n m}(-\mathbf{k}, -\mathbf{k}')\Big]
        \Big[\mathbf{f}_m^*(\mathbf{k}) \cdot \mathbf{T}_{n m}(\mathbf{k}, \mathbf{k}')\Big]
        -
        \Big[\mathbf{f}_n(\mathbf{k}') \cdot \mathbf{T}_{n m}(\mathbf{k}, \mathbf{k}')\Big]
        \Big[\mathbf{T}_{n m}(-\mathbf{k}, -\mathbf{k}') \cdot \mathbf{f}_m^*(\mathbf{k})\Big].
    \end{aligned}
    \end{equation}
    \label{appendixEq:form_factor_expansion}
\end{subequations}
{\color{black}
For systems that admit a mixture of spin singlet and spin triplet correlations from broken parity symmetry, both $C_{\mathrm{sing}-\mathrm{trip}}$ and $C_{\mathrm{trip}-\mathrm{trip}}$ can contribute to the $d$ vector textures.
When tunneling satisfies $T_{nm}(\mathbf{k}, \mathbf{k}') = T_{nm}(-\mathbf{k}, -\mathbf{k}')$, it follows that $C_{\mathrm{sing}-\mathrm{trip}}$ will vanish due to singlet and triplet parts transforming oppositely under parity symmetry, $\mathbf{f}_{n}(\mathbf{k}) = -\mathbf{f}_{n}(-\mathbf{k})$ and $f_{n,0}(\mathbf{k}) = f_{n,0}(-\mathbf{k})$.
For the case of $T_{nm}(\mathbf{k}, \mathbf{k}') \neq T_{nm}(-\mathbf{k}, -\mathbf{k}')$, $C_{\mathrm{sing}-\mathrm{trip}}$ favors antiparallel $d$ vectors parallel to $\hat{\mathbf{n}} = T_{n m; 0}(-\mathbf{k}, -\mathbf{k}') \mathbf{T}_{nm}(\mathbf{k}, \mathbf{k}') - T_{n m; 0}(\mathbf{k}, \mathbf{k}') \mathbf{T}_{nm}(-\mathbf{k}, -\mathbf{k}')$ and will contribute to the Heisenberg-like Josephson coupling between $d$ vectors.
}
In the current work, we exclusively focus on the effective Josephson coupling between spin triplet correlations in adjacent grains; hence, we only consider the last contribution, $C_{\mathrm{trip}-\mathrm{trip}}$.
In the following, we consider the case that 
$\mathbf{T}_{n m} (\mathbf{k}, \mathbf{k}') \neq - \mathbf{T}_{n m}(-\mathbf{k}, -\mathbf{k}')$, which can be realized when tunneling occurs in the presence of a local exchange field breaking time-reversal symmetry, as discussed in the main text.

We now examine a simplified scenario to illustrate the microscopic origin of the effective Josephson coupling between spin triplet superconductors.
Consider the spin triplet pairing correlations and suppose that the spin and orbital degrees are decoupled.
The anomalous Green's function can be decomposed as
\begin{equation}
    \mathcal{F}_m^{(\mathrm{trip})}(\mathbf{k}; i\omega_n)
    =
    \hat{\mathbf{d}}_m
    \cdot
    \boldsymbol{\sigma}
    (i \sigma^y)
    g_m(\mathbf{k}; i\omega_n),
    \label{appendixEq:anomalous_GF_decoupled_spin}
\end{equation}
in which $\mathbf{d}_m$ is a normalized constant $d$ vector that describes the spin part of the triplet pairing correlation and $g_m(\mathbf{k}; i\omega_n)$ encodes the orbital degrees of freedom in the $m^\mathrm{th}$ superconducting grain, satisfying $g_m(\mathbf{k}; i\omega_n) = - g_m(-\mathbf{k}; i\omega_n).$
For simplicity, we consider the case that the tunneling process conserves momentum and satisfies $T_{n m}(\mathbf{k}, \mathbf{k}') = T_{n m} \delta_{\mathbf{k}, \mathbf{k}'}$, with $T_{nm}$ being a constant matrix.
As such, the corresponding Josephson form factor is given by
\begin{equation}
    \mathcal{J}_{nm}^{\mathrm{trip}-\mathrm{trip}}
    =
    w(E_n, E_m; \beta)
    \bigg\{
    \left(T_{nm;0}^2 + \mathbf{T}_{nm}\cdot \mathbf{T}_{nm}\right) 
    \hat{\mathbf{d}}_n\cdot\hat{\mathbf{d}}_m^*
    +
    \left( 2i T_{nm;0} \mathbf{T}_{nm} \right) 
    (\hat{\mathbf{d}}_n \times \hat{\mathbf{d}}_m^*
    )
    -
    \left(\mathbf{T}_{nm} \cdot \hat{\mathbf{d}}_n\right) 
    \left(\mathbf{T}_{nm} \cdot \hat{\mathbf{d}}_m^*\right)
    \bigg\},
    \label{appendixEq:form_factor_w_T}
\end{equation}
in which
\begin{equation}
    w(E_n, E_m; \beta)
    =
    -\frac{1}{\beta}
    \sum_{i\omega_n}
    \sum_{\mathbf{k}}
    g_{n}(\mathbf{k}; i\omega_n)
    g_{m}^*(\mathbf{k}; i\omega_n)
    \label{appendixEq:w_coefficient}
\end{equation}
includes the thermal weighting and is dependent on the BdG quasiparticle energies $E_n$ and $E_m$ at grains $n$ and $m$.
In general, the summation over momentum is nonvanishing when the pairing order parameters of both grains transform under rotation according to the same irrep.
As such, the Josephson form factor between spin triplet superconductors can be compactly expressed as
\begin{equation}
    \mathcal{J}_{n m}^{\mathrm{trip}-\mathrm{trip}}
    =
    2
    \Big(
    J_{n m} \hat{\mathbf{d}}_n \cdot \hat{\mathbf{d}}_m^*
    +
    {\mathbf{D}}_{n m} \cdot
    (\hat{\mathbf{d}}_n \times \hat{\mathbf{d}}_m^* )
    +
    \sum_{a,b}
    \hat{d}_n^{a} {\Gamma}_{n m}^{ab} \hat{d}_m^{b*}
    \Big),
    \label{appendixEq:Josephson_free_energy_spin_triplet}
\end{equation}
in which
\begin{equation}
    \begin{aligned}
        {J}_{n m} &=
        w(E_n, E_m; \beta)
        \left(
        T_{n m; 0}^2
        +
        \mathbf{T}_{n m} \cdot \mathbf{T}_{n m}
        \right),
        \\
        {\mathbf{D}}_{n m} &=
        2i 
        w(E_n, E_m; \beta)
        T_{n m; 0} \mathbf{T}_{n m},
        \\
        {\Gamma}_{n m}^{ab} &= 
        -2 
        w(E_n, E_m; \beta)
        T_{n m}^{a} T_{n m}^b.
    \end{aligned}
    \label{appendixEq:josephson_couplings}
\end{equation}
The effective Josephson couplings ${J}_{n m}$, ${\mathbf{D}}_{n m}$, and ${\Gamma}_{n m}$ correspond to the Heisenberg-like term, Dzyaloshinskii-Moriya(DM)-like antisymmetric term, and symmetric anisotropic coupling term respectively, as discussed in Eq.~\eqref{josephson_coupling.d_vectors} in the main text.
It should be noted that, here, $T_{nm}$ corresponds to the microscopic tunneling processes across the entire interface between grains $n$ and $m$.

We emphasize that to realize the DM-like antisymmetric coupling in Eq.~\eqref{appendixEq:form_factor_expansion}, it is {necessary and sufficient} for the effective tunneling to satisfy
\begin{equation}
    T_{nm;0}(\mathbf{k}, \mathbf{k}')
    \mathbf{T}_{nm}(-\mathbf{k}, -\mathbf{k}')
    \neq
    - T_{nm;0}(-\mathbf{k}, -\mathbf{k}')
    \mathbf{T}_{nm}(\mathbf{k}, \mathbf{k}').
    \label{appendixEq:dm_coupling_condition}
\end{equation}
Hence, it is {sufficient} for the spin-independent and spin-dependent tunneling $T_{nm;0}(\mathbf{k}, \mathbf{k}')$ and $\mathbf{T}_{nm}(\mathbf{k}, \mathbf{k}')$ to be even functions of $\mathbf{k}$ and $\mathbf{k}'$ for the DM-like Josephson coupling to be nonvanishing.
The antisymmetric Josephson coupling cannot arise, for example, through spin-orbit coupling alone, under which $\mathbf{T}_{nm}(\mathbf{k}, \mathbf{k}') = -\mathbf{T}_{nm}(-\mathbf{k}, -\mathbf{k}')$ for real-valued $\mathbf{T}_{nm}$.
Rather, the tunneling requires, for example, the presence of a local exchange field or impurities at the interface that break time reversal symmetry.
Lastly, if the spin and orbital degrees of freedom of the anomalous Green's function cannot be decoupled, it is necessary to consider the summation over the orbital degrees of freedom.
Nonetheless, the key qualitative features, such as the anisotropic Josephson exchange, remain robust and can be captured phenomenologically.

\subsection{Coefficients in Josephson couplings}
We derive the coefficients in the Josephson couplings from the form factor in Eq.~\eqref{appendixEq:form_factor_w_T} for a simple representative scenario.
Suppose the pairing correlator takes the form
\begin{equation}
    \mathcal{F}_{m}^{(\mathrm{trip})}(\mathbf{k}; i\omega_n) = - \frac{\Delta_0}{\omega_n^2 + E_m^2} \tilde{g}_m(\mathbf{k}) \hat{\mathbf{d}}_m \cdot \boldsymbol{\sigma} (i\sigma^y),
    \label{appendixEq:anomalous_GF_simplified_scenario}
\end{equation}
in which $\hat{\mathbf{d}}_m$ is the $d$ vector, and $\tilde{g}_m(\mathbf{k})$ describes the orbital degree of freedom.
Here, $E_m$ is the BdG quasiparticle energy for the $m^{\mathrm{th}}$ grain and $\Delta_0 
$ the magnitude of the pairing gap.
In general, the anomalous Green's function is obtained by solving Gor'kov's equations and depends on the pairing gap function and electronic band structure, as detailed in Sec.~\ref{SM:pairing_correlations}.
For the form of the anomalous Green's function in Eq.~\eqref{appendixEq:anomalous_GF_simplified_scenario}, the Matsubara frequency summation in Eq.~\eqref{appendixEq:w_coefficient} is given by
\begin{align}
    - \frac{1}{\beta}\sum_{i\omega_n}
    \frac{1}{\omega_n^2 + E_n^2}
    \frac{1}{\omega_n^2 + E_m^2}
    &=
    \frac{1}{2E_n E_m}
    \bigg\{
    \frac{n_F(E_n) + n_F(E_m) - 1}{E_n+ E_m}
    -
    \frac{n_F(E_n) - n_F(E_m)}{E_n - E_m}
    \bigg\}.
    \label{matsubara_frequency_summation_example_case}
\end{align}
Above,
\begin{math}
    n_F(E) = (1 + e^{\beta E})^{-1}
\end{math}
is the Fermi-Dirac distribution, and we have used
\begin{math}
    n_F(-E) = 
    1- n_F(E).
\end{math}
It follows that
\begin{equation}
    w(E_n, E_m; \beta) = 
    \Delta_0^2
    \sum_{\mathbf{k}}
    \frac{1}{2E_n E_m}
    \bigg\{
    \frac{n_F(E_n) + n_F(E_m) - 1}{E_n+ E_m}
    -
    \frac{n_F(E_n) - n_F(E_m)}{E_n - E_m}
    \bigg\}
    \tilde{g}_n(\mathbf{k})
    \tilde{g}_m^*(\mathbf{k}).
\end{equation}
At zero-temperature, the above equation reduces to
\begin{equation}
    w(E_n, E_m; \beta\rightarrow \infty) = 
    -\Delta_0^2
    \sum_{\mathbf{k}}
    \frac{1}{2E_n E_m}
    \frac{\tilde{g}_n(\mathbf{k})
    \tilde{g}_m^*(\mathbf{k})}{E_n+ E_m}
    .
\end{equation}
{\color{black} In the weak-coupling regime, states at the Fermi surface primarily contribute to Josephson tunneling, with density of states $D(0) \approx 1/W$, in which $W$ is the bandwidth.
The weighting factor is given by~\cite{Frazier2025b}
\begin{equation}
    w_{nm} \approx
    - \frac{\Delta_0 D^2(0)}{2} I_{nm}
\end{equation}
up to an overall overlap of $\tilde{g}_n(\mathbf{k})$ and $\tilde{g}_m(\mathbf{k})$ in $I_{nm}$, which is of order unity.
Conseqeuntly, the weighting factor scales as
\begin{equation}
    w_{nm} \sim -D^2(0) \Delta_0 \sim - \frac{\Delta_0}{W^2},
\end{equation}
and the magnitude of Josephson tunneling is on the order of the amplitude of the pairing gap.
}

\subsection{Josephson couplings for momentum-dependent \texorpdfstring{$d$}{d} vector}

We now consider the case in which spin and momentum degrees are coupled, for instance due to strong spin-orbit interaction.
In this situation, the anomalous Green's function cannot be written in the simplified decoupled form in 
Eq.~\eqref{appendixEq:anomalous_GF_decoupled_spin}, and one must account for a momentum-dependent $d$ vector.

For simplicity, suppose the system has rotational symmetry and that the anomalous Green's function can be written as
\begin{equation}
    \mathcal{F}^{(\mathrm{trip})}_m(\mathbf{k}; i\omega_n)
    =
    - \frac{\Delta_0}{\omega_n^2 + E_{m, \mathbf{k}}^2}
    K_m(k)
    \hat{\mathbf{d}}_m(\hat{\mathbf{k}})\cdot \boldsymbol{\sigma} (i\sigma_y).
\end{equation}
Here, $E_{m, \mathbf{k}}$ is the BdG quasiparticle energy, $K_m(k)$ is a scalar function of $|\mathbf{k}|$, and 
$\hat{\mathbf{d}}_m(\hat{\mathbf{k}})$ is a momentum-dependent $d$ vector satisfying $\hat{\mathbf{d}}_m(\hat{\mathbf{k}}) = - \hat{\mathbf{d}}_m(-\hat{\mathbf{k}})$.
Inserting this into Eq.~\eqref{appendixEq:form_factor_expansion}, the resulting form factor describing Josephson coupling between two spin triplet superconducting grains reduces to
\begin{equation}
    \mathcal{J}_{nm}^{\mathrm{trip}-\mathrm{trip}}
    =
    2 \sum_{\mathbf{k}, \mathbf{k}'}
    \Big(
    J_{n m}(\mathbf{k}, \mathbf{k}') \hat{\mathbf{d}}_n (\mathbf{k}) \cdot \hat{\mathbf{d}}_m^* (\mathbf{k}') 
    +
    {\mathbf{D}}_{n m}(\mathbf{k}, \mathbf{k}') \cdot
    (\hat{\mathbf{d}}_n (\mathbf{k}) \times \hat{\mathbf{d}}_m^* (\mathbf{k}')  )
    +
    \sum_{a,b}
    \hat{d}_n^{a} (\mathbf{k}) {\Gamma}_{n m}^{ab}(\mathbf{k}, \mathbf{k}') \hat{d}_m^{b*} (\mathbf{k}') 
    \Big),
    \label{appendixEq:josephson_coupling_k-dependence}
\end{equation}
in which the three Josephson coupling amplitudes are given by
\begin{subequations}
    \begin{align}
        J_{nm}(\mathbf{k}, \mathbf{k}')
        &=
        u(E_{n, \mathbf{k}}, E_{m, \mathbf{k}'}; \beta) 
        K_n(k) K^*_m(k')
        \Big(T_{nm; 0}(\mathbf{k}, \mathbf{k}')
        T_{nm; 0}(-\mathbf{k}, -\mathbf{k}')
        +
        \mathbf{T}_{nm}(\mathbf{k}, \mathbf{k}') \cdot \mathbf{T}_{nm}(-\mathbf{k}, -\mathbf{k}')
        \Big),
        \\
        \mathbf{D}_{nm}(\mathbf{k}, \mathbf{k}')
        &=
        i u(E_{n, \mathbf{k}}, E_{m, \mathbf{k}'}; \beta) K_n(k) K^*_m(k')
        \Big(T_{nm;0}(\mathbf{k}, \mathbf{k}')\mathbf{T}_{nm}(-\mathbf{k}, -\mathbf{k}')
        +
        T_{nm;0}(-\mathbf{k}, -\mathbf{k}')\mathbf{T}_{nm}(\mathbf{k}, \mathbf{k}')
        \Big),
        \\
        \Gamma^{ab}_{nm} (\mathbf{k}, \mathbf{k}')
        &=
        - u(E_{n, \mathbf{k}}, E_{m, \mathbf{k}'}; \beta) K_n(k) K^*_m(k')
        \Big(\mathbf{T}_{nm}^a(\mathbf{k}, \mathbf{k}')
        \mathbf{T}_{nm}^b(-\mathbf{k}, -\mathbf{k}')
        + \mathbf{T}_{nm}^a(-\mathbf{k}, -\mathbf{k}')
        \mathbf{T}_{nm}^b(\mathbf{k}, \mathbf{k}')
        \Big).
    \end{align}
\end{subequations}
Above,
\begin{math}
    u(E_{n, \mathbf{k}}, E_{m, \mathbf{k}'}; \beta) = \frac{\Delta_0^2}{2E_{n, \mathbf{k}} E_{m, \mathbf{k}'}}
    \{
    \frac{n_F(E_{n, \mathbf{k}}) + n_F(E_{m, \mathbf{k}'}) - 1}{E_{n, \mathbf{k}}+ E_{m, \mathbf{k}'}}
    -
    \frac{n_F(E_{n, \mathbf{k}}) - n_F(E_{m, \mathbf{k}'})}{E_{n, \mathbf{k}} - E_{m, \mathbf{k}'}}
    \}.
\end{math}

To isolate the angular dependence of the $d$ vector, we suppose that the tunneling has the form $T_{nm}(\mathbf{k}, \mathbf{k}') = T_{nm} \delta_{\mathbf{k}, \mathbf{k}'}$.
Under this simplification, the Heisenberg-like, DM-like, and $\Gamma$-type couplings contributing to the Josephson form factor can respectively be expressed as
\begin{subequations}
    \begin{align}
        \mathcal{J}_{nm}^{J}
        &=
        2 \frac{A}{(2\pi)^2}
        (T_{nm;0}^2 + \mathbf{T}_{nm} \cdot \mathbf{T}_{nm})
        \int \mathrm{d}^2 k \,
        u(E_{n, \mathbf{k}}, E_{m, \mathbf{k}}; \beta)
        K_n(k)K_m^*(k)
        \hat{\mathbf{d}}_n(\hat{\mathbf{k}}) \cdot \hat{\mathbf{d}}_m^*(\hat{\mathbf{k}}) ,
        \\
        \mathcal{J}_{nm}^{\mathrm{DM}}
        &=
        4i \frac{A}{(2\pi)^2}
        (T_{nm;0} \mathbf{T}_{nm})
        \int \mathrm{d}^2 k \,
        u(E_{n, \mathbf{k}}, E_{m, \mathbf{k}}; \beta)
        K_n(k)K_m^*(k)
        \hat{\mathbf{d}}_n(\hat{\mathbf{k}}) \times \hat{\mathbf{d}}_m^*(\hat{\mathbf{k}}),
        \\
        \mathcal{J}_{nm}^\Gamma
        &=
        -4 \frac{A}{(2\pi)^2}
        {T}^a_{nm} {T}^b_{nm}
        \int \mathrm{d}^2 k \,
        u(E_{n, \mathbf{k}}, E_{m, \mathbf{k}}; \beta)
        K_n(k)K_m^*(k)
        \hat{{d}}_n^a(\hat{\mathbf{k}}) \hat{{d}}_m^{b*}(\hat{\mathbf{k}}),
    \end{align}
\end{subequations}
where $A$ is the area of the two-dimensional superconducting grain.
Because $\hat{\mathbf{d}}(\hat{\mathbf{k}})$ is an odd function of $\hat{\mathbf{k}}$, the integrands above are generally even in $\mathbf{k}$ and can yield nonvanishing contributions.
It follows that the anisotropic Josephson couplings presented in Eq.~\eqref{josephson_coupling.d_vectors} in the main text persist even in system with, for example, spin-orbit coupling.
As such, minimizing the free energy can still result in a spatially inhomogeneous $d$ vector texture.


\subsection{Josephson free energy}

We review the Josephson free energy and relate it to the Josephson form factor.
From the Ambegaokar-Baratoff formalism, the Josephson tunneling current at zero voltage bias between two grains labelled by $n$ and $m$ is given by~\cite{Ambegaokar1963, Sigrist1991, Frazier2024}
\begin{equation}
    I_J^{(n,m)} = i \frac{e}{\hbar}\Big(
   \mathcal{J}_{n m} 
    - \mathrm{c.c.}
    \Big),
    \label{appendixEq:josephson_current}
\end{equation}
with $\mathcal{J}_{n m}$ being the Josephson form factor in Eq.~\eqref{form_factor} of the main text.
Explicitly separating the $\mathrm{U}(1)$ phase from the superconducting pairing correlations, it follows that
\begin{math}
    I_J^{(n,m)} = i \frac{e}{\hbar}(
    \mathcal{J}_{n m}
    e^{i\Delta \phi_{n m}} 
    - \mathrm{c.c.}
    ),
\end{math}
with $\Delta \phi_{n m} = \phi_n - \phi_m$ being the difference in the $\mathrm{U}(1)$ phases of the pair condensate at the two grains.

To compare, the first order Josephson coupling in Ginzburg-Landau theory between the $n^\mathrm{th}$ and $m^\mathrm{th}$ grains is given by 
\begin{equation}
    F_{nm}[\Delta \phi_{n m}] = 
    F_J e^{i\Delta \phi_{n m}} + \mathrm{c.c.}
\end{equation}
in which $F_J$ is the real-valued Josephson coupling amplitude.
The corresponding Josephson current is given by
\begin{equation}
    I_J^{(n,m)} = \frac{2e}{\hbar} \frac{\partial F_{nm}}{\partial \Delta \phi_{n m}}
    =
    i \frac{2e}{\hbar}
    \Big(
    F_J e^{i\Delta \phi_{nm}}
    -
    \mathrm{c.c.}
    \Big)
\end{equation}
Comparison to the Josephson current in Eq.~\eqref{appendixEq:josephson_current} shows that the Josephson free energy can be expressed in terms of the Josephson form factor, 
\begin{equation}
    F_{nm} 
    \sim
    \frac{1}{2}
    \left( \mathcal{J}_{n m} + \mathrm{c.c.} \right).
    \label{appendixEq:josephson_free_energy_general}
\end{equation}
The relation to the Josephson form factor reflects the fact that the form factor describes the microscopic tunneling processes that underlie the Josephson coupling between grains.
Hence, the Josephson form factor naturally appears in the free energy describing spatial variations of the superconducting order parameter.

%% file: supplemental/s2.model.tex
\section{Band Hamiltonians}

\label{SM:sd_models}

In this section, we provide the tight-binding models for the three-sublattice systems on the triangular and kagome lattices used in analyzing the superconducting correlations and effective tunneling.
For both systems, we are considering a local spin texture formed by three spins labelled $\mathbf{s}_a$, $\mathbf{s}_b$, and $\mathbf{s}_c$, as shown in Fig.~\ref{fig:triangular+kagome_lattice}(a).
The Hamiltonian for the system can be written generally as
\begin{equation}
    H
    = H_\mathrm{kin} + H_{sd}
\end{equation}
in which $H_\mathrm{kin}$ and $H_{sd}$ describe nearest neighbor hopping and local $s$-$d$ exchange.
The kinetic Hamiltonian is given by
\begin{equation}
    H_{\mathrm{kin}} = t_0 \sum_{\langle \mathbf{r}_i, \mathbf{r}_j \rangle, \alpha}
    c^\dagger_{\mathbf{r}_i, \alpha}
    c_{\mathbf{r}_j, \alpha}
    -\mu \sum_{\mathbf{r}_i, \alpha} c^\dagger_{\mathbf{r}_i, \alpha} c_{\mathbf{r}_i, \alpha},
    \label{appendixEq:kinetic.NN.real_space}
\end{equation}
and the $s$-$d$ coupling, at the mean-field level, is given by
\begin{equation}
    H_{sd} = 
    J_{sd}
    \sum_{\mathbf{r}_j, \alpha, \alpha'}
    c^{\dagger}_{\mathbf{r}_j, \alpha}
    [
    \boldsymbol{\sigma}_{\alpha, \alpha'}
    \cdot
    \mathbf{s}_j
    ]
    c_{\mathbf{r}_j, \alpha'},
    \label{appendixEq:sd_coupling_ham.real_space}
\end{equation}
as described in the main text.

\subsection{Kagome lattice}

Consider the kagome lattice, shown schematically in Fig.~\ref{fig:triangular+kagome_lattice}(b).
We denote nearest neighbor vectors as
\begin{gather}
    \boldsymbol{\delta}_1 = \frac{a}{2} (1, 0)^\mathrm{T}; 
    \hspace{2em}
    \boldsymbol{\delta}_2 = \frac{a}{2} \left( \frac{1}{2}, \frac{\sqrt{3}}{2}\right)^\mathrm{T}; 
    \hspace{2em}
    \boldsymbol{\delta}_3 = \frac{a}{2} \left( -\frac{1}{2}, \frac{\sqrt{3}}{2}\right)^\mathrm{T},
\end{gather}
and lattice vectors as
\begin{gather}
    \mathbf{a}_1 =  a (1, 0)^\mathrm{T};
    \hspace{2em}
    \mathbf{a}_2 
    = 
    a \left( \frac{1}{2}, \frac{\sqrt{3}}{2}\right)^\mathrm{T}.
\end{gather}
The reciprocal lattice vectors satisfying $\mathbf{a}_i \cdot \mathbf{b}_j = 2\pi \delta_{ij}$ are given by
\begin{math}
    \mathbf{b}_1 = b \left( \frac{\sqrt{3}}{2}, - \frac{1}{2} \right)^\mathrm{T}
\end{math}
and
\begin{math}
    \mathbf{b}_2 = b (0, 1)^\mathrm{T},
\end{math}
with $b = 4\pi/ \sqrt{3} a$.

We now consider the Fourier transform of the nearest-neighbor tight binding model in Eq.~\eqref{appendixEq:kinetic.NN.real_space}.
In the $(\mathbf{c}_{a, \mathbf{k}}, \mathbf{c}_{b, \mathbf{k}}, \mathbf{c}_{c, \mathbf{k}})^\mathrm{T}$ basis, the kinetic Hamiltonian is given by
\begin{equation}
    \mathcal{H}_\mathrm{kin}(\mathbf{k})
    =
    -
    \mu \mathbbm{1}_3 \otimes \sigma^0
    + 2 t_0 \left(
    \begin{array}{ccc}
         0 & \cos \alpha_1  & \cos \alpha_2
         \\
         \cos \alpha_1 & 0 & \cos \alpha_3
         \\
         \cos \alpha_2 & \cos \alpha_3 & 0
    \end{array}
    \right)
    \otimes \sigma^0,
    \label{kagome.kinetic}
\end{equation}
in which $\alpha_i \equiv \mathbf{k} \cdot \boldsymbol{\delta}_{i},$ with $i = 1,2,3$.
This gives rise to two dispersive bands, and a flat band from the destructive interference, characteristic of the kagome geometry.
The coupling to the local spins of the $d$ electrons is purely on-site and thus has no momentum dependence.
The $s$-$d$ exchange is given by 
\begin{equation}
    \mathcal{H}_{sd} = 
    J_{sd}
    \left(
    \begin{array}{ccc}
         \mathbf{s}_a \cdot \boldsymbol{\sigma} & 0 & 0
         \\
         0 & \mathbf{s}_b \cdot \boldsymbol{\sigma} & 0
         \\
         0 & 0 & \mathbf{s}_c \cdot \boldsymbol{\sigma}
    \end{array}
    \right),
    \label{appendixEq:sd_term}
\end{equation}
and is diagonal in sublattice space.

\subsection{Triangular lattice}
\begin{figure}
    \centering
    \includegraphics[width=\linewidth]{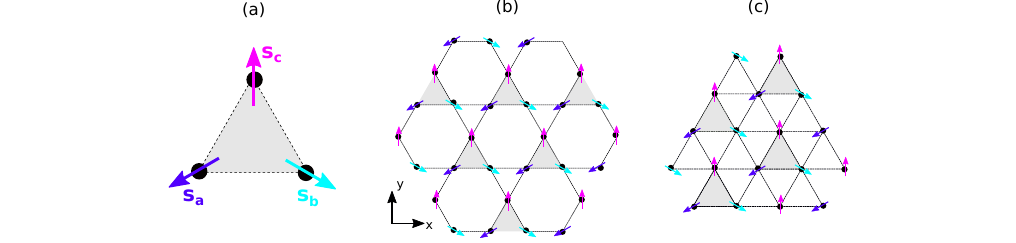}
    \caption{(a)~Three spins $\mathbf{s}_a$, $\mathbf{s}_b$, and $\mathbf{s}_c$ comprising the three sublattices in purple, cyan, and pink, respectively.
    An example coplanar $120^\circ$ ordered spin configuration is shown.
    (b)~Triangular lattice and (c)~kagome lattice are shown with the corresponding three sublattices highlighted in gray.}
    \label{fig:triangular+kagome_lattice}
\end{figure}

Now, consider a triangular lattice, as shown in Fig.~\ref{fig:triangular+kagome_lattice}(c).
Nearest neighbors are related by
\begin{equation}
    \boldsymbol{\delta}_1 = \frac{a}{\sqrt{3}} (1, 0)^\mathrm{T}; 
    \hspace{2em}
    \boldsymbol{\delta}_2 = \frac{ a}{\sqrt{3}} \left( \frac{1}{2}, \frac{\sqrt{3}}{2}\right)^\mathrm{T}; 
    \hspace{2em}
    \boldsymbol{\delta}_3 = \frac{a}{\sqrt{3}} \left( -\frac{1}{2}, \frac{\sqrt{3}}{2}\right)^\mathrm{T}.
    \label{appendixEq:nearest_neighbour_triangular}
\end{equation}
Due to the three-sublattice magnetic ordering, the unit cell is enlarged, with lattice vectors given by
\begin{equation}
    \mathbf{a}_1 =  
    a \left( \frac{\sqrt{3}}{2}, \frac{1}{2} \right)^\mathrm{T};
    \hspace{2em}
    \mathbf{a}_2 
    = a \left( 0, 1\right)^\mathrm{T};
    \hspace{2em}
    \mathbf{a}_3 = \mathbf{a}_2 - \mathbf{a}_1 = a \left( - \frac{\sqrt{3}}{2}, \frac{1}{2} \right)^\mathrm{T}.
\end{equation}
Here, $a = \sqrt{3} \delta$.
Reciprocal lattice vectors  are given by
\begin{math}
    \mathbf{b}_1 = b (1, 0)^\mathrm{T}
\end{math}
and
\begin{math}
    \mathbf{b}_2 = b \left(- \frac{1}{2},  \frac{\sqrt{3}}{2} \right)^\mathrm{T},
\end{math}
with $b = 4\pi/ \sqrt{3} a$.
In the $(\mathbf{c}_{a, \mathbf{k}}, \mathbf{c}_{b, \mathbf{k}}, \mathbf{c}_{c, \mathbf{k}})^\mathrm{T}$ basis, the kinetic Hamiltonian is given by
\begin{equation}
    \mathcal{H}_\mathrm{kin}(\mathbf{k})
    =
    -\mu \mathbbm{1}_3 \otimes \sigma^0
    + t_0
    \left(
    \begin{array}{ccc}
        0 & 
        \left( e^{-i \mathbf{k} \cdot \boldsymbol{\delta}_1} + e^{-i \mathbf{k} \cdot \boldsymbol{\delta}_3} + e^{+i \mathbf{k} \cdot \boldsymbol{\delta}_2}\right)
        &
        \left( e^{-i \mathbf{k} \cdot \boldsymbol{\delta}_2} + e^{+i \mathbf{k} \cdot \boldsymbol{\delta}_1} + e^{+i \mathbf{k} \cdot \boldsymbol{\delta}_3}\right)
        \\
        \mathrm{h.c.} & 0 & \left( e^{-i \mathbf{k} \cdot \boldsymbol{\delta}_1} + e^{-i \mathbf{k} \cdot \boldsymbol{\delta}_3} + e^{+i \mathbf{k} \cdot \boldsymbol{\delta}_2}\right)
        \\
        \mathrm{h.c.} & \mathrm{h.c.} & 0
    \end{array}
    \right) \otimes \sigma^0.
    \label{appendixEq:triangular_lattice.kinetic_ham}
\end{equation}
Lastly, the $s$-$d$ term is the same as that in Eq.~\eqref{appendixEq:sd_term},
which is diagonal in sublattice space.

Furthermore, we consider the case of Ising-type spin-orbit coupling on the triangular lattice.
In real space, the Ising spin-orbit coupling has the form
\begin{equation}
    \begin{aligned}
        &[\mathcal{H}_\mathrm{SO}]_{\mathbf{r}_n, \mathbf{r}_{n'}}
        =
        \lambda_\mathrm{z} \left(
        \begin{array}{ccc}
            0 &  
            i(\delta_{\mathbf{r}_n, (\mathbf{r}_{n'} + \boldsymbol{\delta}_1)} +
            \delta_{\mathbf{r}_n, (\mathbf{r}_{n'} + \boldsymbol{\delta}_3)} +
            \delta_{\mathbf{r}_n, (\mathbf{r}_{n'} - \boldsymbol{\delta}_2)})
            & 
            -i(\delta_{\mathbf{r}_n, (\mathbf{r}_{n'} + \boldsymbol{\delta}_2)} +
            \delta_{\mathbf{r}_n, (\mathbf{r}_{n'} - \boldsymbol{\delta}_1)} +
            \delta_{\mathbf{r}_n, (\mathbf{r}_{n'} - \boldsymbol{\delta}_3)})
            \\
            \mathrm{h.c.} & 0 & 
            i(\delta_{\mathbf{r}_n, (\mathbf{r}_{n'} + \boldsymbol{\delta}_1)} +
            \delta_{\mathbf{r}_n, (\mathbf{r}_{n'} + \boldsymbol{\delta}_3)} +
            \delta_{\mathbf{r}_n, (\mathbf{r}_{n'} - \boldsymbol{\delta}_2)})
            \\
            \mathrm{h.c.} & \mathrm{h.c.} & 0
        \end{array}
        \right)
        \otimes \sigma^z,
    \end{aligned}
\end{equation}
in which $\lambda_z$ is the strength of the spin-orbit coupling.
Its Fourier transform is given by
\begin{equation}
    \mathcal{H}_\mathrm{SO}(\mathbf{k})
    =
    \lambda_z
    \left(
    \begin{array}{ccc}
        0 & 
        i\left( e^{-i \mathbf{k} \cdot \boldsymbol{\delta}_1} + e^{-i \mathbf{k} \cdot \boldsymbol{\delta}_3} + e^{+i \mathbf{k} \cdot \boldsymbol{\delta}_2}\right)
        &
        -i \left( e^{-i \mathbf{k} \cdot \boldsymbol{\delta}_2} + e^{+i \mathbf{k} \cdot \boldsymbol{\delta}_1} + e^{+i \mathbf{k} \cdot \boldsymbol{\delta}_3}\right)
        \\
        \mathrm{h.c.} & 0 & i\left( e^{-i \mathbf{k} \cdot \boldsymbol{\delta}_1} + e^{-i \mathbf{k} \cdot \boldsymbol{\delta}_3} + e^{+i \mathbf{k} \cdot \boldsymbol{\delta}_2}\right)
        \\
        \mathrm{h.c.} & \mathrm{h.c.} & 0
    \end{array}
    \right)\otimes \sigma^z.
    \label{appendixEq:triangular_lattice.Ising_SO}
\end{equation}
Such Ising spin-orbit coupling can lead to spin-valley polarization, for example, in TaS$_2$.

%% file: supplemental/s3.pairing_correlations.tex
\section{Superconducting pairing correlations}

\label{SM:pairing_correlations}


In this section, we derive the pairing correlations in the $s$-$d$ model on the three-sublattice system.
Consider the Bogoliubov-De Gennes (BdG) Hamiltonian in the Nambu basis $(\mathbf{c}_{\mathbf{k}, \uparrow}, \mathbf{c}_{\mathbf{k}, \downarrow}, \mathbf{c}^\dagger_{-\mathbf{k}, \uparrow}, \mathbf{c}^\dagger_{-\mathbf{k}, \downarrow})^\mathrm{T},$ given by
\begin{equation}
    \mathcal{H}_\mathrm{BdG}(\mathbf{k})
    =
    \left(
    \begin{array}{cc}
         \mathcal{H}
         (\mathbf{k})
         & \Delta(\mathbf{k})
         \\
         \Delta^\dagger(\mathbf{k})
         &
         - \mathcal{H}^\mathrm{T}(-\mathbf{k})
    \end{array}
    \right)
    \label{AppendixEq:BdG_Ham}
\end{equation}
Here, $\Delta(\mathbf{k})$ is the pairing gap function.
The Nambu-Gor'kov Green's function is defined as 
\begin{equation}
    \mathcal{G}_\mathrm{BdG} (\mathbf{k}; \tau)
    =
    - \left \langle 
    \mathcal{T}_\tau
    \left(
    \begin{array}{c}
         \mathbf{c}_{\mathbf{k}}(\tau) \\ \mathbf{c}^\dagger_{-\mathbf{k}} (\tau)
    \end{array}
    \right)
    \left(
    \begin{array}{c}
         \mathbf{c}^\dagger_{\mathbf{k}}(0), \mathbf{c}_{-\mathbf{k}} (0)
    \end{array}
    \right)
    \right \rangle
    =
    \left(
    \begin{array}{cc}
         \mathcal{G}_{\mathrm{BdG}; 1,1}(\mathbf{k}; \tau)
         &\mathcal{F}(\mathbf{k}; \tau)
         \\
         \mathcal{F}^\dagger(\mathbf{k}; \tau)
         &
         \mathcal{G}_{\mathrm{BdG}; 2,2}(\mathbf{k}; \tau)
    \end{array}   
    \right).
\end{equation}
Here, the off-diagonal anomalous Green's function, defined as
\begin{math}
    \mathcal{F}(\mathbf{k}; \tau) = - \langle \mathcal{T}_\tau \mathbf{c}_\mathbf{k}(\tau) \mathbf{c}_{-\mathbf{k}}(0) \rangle,
\end{math}
describes the pairing correlations.

Due to the coupling to the exchange field formed by local moments of the $d$ electrons, spin triplet pairing correlations naturally arise.
The anomalous Green's function, $\mathcal{F}_{i, \alpha; j, \beta} (\mathbf{k}; i\omega) = - \int \mathrm{d} \tau  e^{i\omega \tau} \langle \mathcal{T}_\tau c_{-\mathbf{k}, j , \beta} (\tau) c_{\mathbf{k}, i, \alpha}(0) \rangle$, is obtained from solving the Gor'kov equations and is given by the off-diagonal component of the BdG Green's function $\mathcal{G}_\mathrm{BdG}(\mathbf{k}; i\omega) = (i\omega - \mathcal{H}_\mathrm{BdG}(\mathbf{k}))^{-1}$, which corresponds to the pairing correlators~\cite{Abrikosov1963, Lueders1971}.
The anomalous Green's function in frequency space is in general given by~\cite{Lueders1971, Frazier2024}
\begin{equation}
    \mathcal{F}(\mathbf{k}; i\omega) = - \mathcal{G}
    (\mathbf{k}; i\omega)
    \Delta(\mathbf{k})
    \Big[
    \mathbbm{1} + \mathcal{G}^\mathrm{T}(-\mathbf{k}; -i\omega)
    \Delta^\dagger (\mathbf{k}) \mathcal{G}
    (\mathbf{k}; i\omega) \Delta(\mathbf{k}) \Big]^{-1}
    \mathcal{G}^\mathrm{T}(-\mathbf{k}; -i\omega),
    \label{anomalous_GF.freq_space}
\end{equation}
in which 
\begin{math}
    \mathcal{G}
    (\mathbf{k}; i\omega) 
    =
    (i\omega - \mathcal{H}
    (\mathbf{k}))^{-1}
\end{math}
is the single-particle Green's function, including the 
$s$-$d$ exchange.
In the linearized gap regime, the anomalous Green's function is approximated by
\begin{equation}
    \mathcal{F}(\mathbf{k}; i\omega) \approx 
    - \mathcal{G}
    (\mathbf{k}; i\omega) \Delta(\mathbf{k}) \mathcal{G}^\mathrm{T}(-\mathbf{k}; -i\omega).
    \label{appendixEq:anomalous_GF.linearized_gap_regime}
\end{equation}
{\color{black}
The approximation is valid when the order of magnitude of $\mathcal{G}^\mathrm{T}(-\mathbf{k}; -i\omega)
\Delta^\dagger (\mathbf{k}) \mathcal{G}
(\mathbf{k}; i\omega) \Delta(\mathbf{k}) \ll 1$.

When $J_{sd} < t_0$, we can expand $\mathcal{G}(\mathbf{k}; i\omega)$ in orders of $J_{sd}/t_0$ as}
\begin{equation}
    \mathcal{G}
    (\mathbf{k}; i\omega)
    =
    \mathcal{G}_\mathrm{kin}(\mathbf{k}; i\omega)
    \sum_{n \geq 0}
    [\mathcal{H}_{sd} \mathcal{G}_\mathrm{kin}(\mathbf{k}; i\omega)]^n,
    \label{Gband.dyson_expansion}
\end{equation}
with $\mathcal{G}_\mathrm{kin}(\mathbf{k}; i\omega) = (i\omega - \mathcal{H}_\mathrm{kin}(\mathbf{k}))^{-1}$ being the free Green's function in the absence of the $s$-$d$ exchange.
{\color{black}The order of magnitude of $\mathcal{G}^\mathrm{T}(-\mathbf{k}; -i\omega)
\Delta^\dagger (\mathbf{k}) \mathcal{G}
(\mathbf{k}; i\omega) \Delta(\mathbf{k})$ at any $N\mathrm{th}$ order of $J_{sd}/t_0$ becomes $(\Delta_0/t_0)^2 (J_{sd}/t_0)^N$, which is $\ll 1$ for $\Delta_0/t_0 \ll 1$.
Alternatively, when $J_{sd} > t_0$, we can expand $\mathcal{G}(\mathbf{k}; i\omega)$ in orders of $t_0/J_{sd}$, and the order of magnitude of $\mathcal{G}^\mathrm{T}(-\mathbf{k}; -i\omega)
\Delta^\dagger (\mathbf{k}) \mathcal{G}
(\mathbf{k}; i\omega) \Delta(\mathbf{k})$ at any $N\mathrm{th}$ order of $t_0/J_{sd}$ becomes $(\Delta_0/J_{sd})^2 (t_0/J_{sd})^N$, which is $\ll 1$ for $\Delta_0/J_{sd} \ll 1$.
}

For $J_{sd} < t_0$, the anomalous Green's function
can be systematically expanded in orders of $J_{sd}$ as $\mathcal{F}(\mathbf{k}; i\omega) = \sum_N \mathcal{O}(J_{sd}^N)$, in which the $N\mathrm{th}$ order contribution is given by
\begin{align}
    & 
    \mathcal{O}(J_{sd}^{N})
    = 
    \Delta_0
    \sum_{n = 0}^{N}
    (-1)^{N-n+1}
    \mathcal{G}_\mathrm{kin}(\mathbf{k}; i\omega) 
    \Big[
    \mathcal{H}_{sd} \mathcal{G}_\mathrm{kin}(\mathbf{k}; i\omega)
    \Big]^n
    \mathcal{G}_\mathrm{kin}^\mathrm{T}(-\mathbf{k}; -i\omega)
    \Big[
    \mathcal{H}_{sd} \mathcal{G}_\mathrm{kin}^\mathrm{T}(-\mathbf{k}; -i\omega)
    \Big]^{N-n}
    (\mathbbm{1}_{3 \times 3} \otimes i\sigma^y).
\end{align}
Above, $\mathcal{G}_\mathrm{kin}$ has off-diagonal components corresponding to inter-sublattice hopping while $\mathcal{H}_{sd}$ introduces spin-sublattice entanglement via the local exchange field.
In particular, the noncollinear spin texture leads to emergence of mixed-parity superconducting correlations in the spin singlet and spin triplet channels.
{\color{black}Here, we have retained all terms in the expansion with respect to $J_{sd}$.
Generally, we consider the case in which the spin-texture of the bands at the Fermi surface is determined by the $s$-$d$ exchange, with $\Delta_0 < J_{sd}$.
}

To demonstrate the singlet and triplet pairing structure explicitly,  
we decompose the anomalous Green's function between sublattices $i$ and $j$ in Eq.~\eqref{appendixEq:anomalous_GF.linearized_gap_regime} into the spin basis as
\begin{equation}
    \mathcal{F}_{ij}(\mathbf{k}; i\omega) = 
    \Big[
    f_{0; ij}(\mathbf{k}; i\omega)
    +
    \mathbf{f}_{ij}(\mathbf{k}; i\omega) \cdot \boldsymbol{\sigma} 
    \Big]
    (i\sigma^y),
    \label{singlet_triplet_pairing_correlations}
\end{equation}
in which $f_{0; ij}$ and $\mathbf{f}_{ij}$ denote the singlet and triplet pairing correlations respectively.
Here, $\mathbf{f}_{ij}$ plays a role analogous to the $\mathbf{d}$ vector in Ginzburg-Landau theory, corresponding to induced spin triplet correlations.
In general, noncollinear spin configurations can lead to finite triplet superconducting correlations, with the direction and magnitude of $\mathbf{f}_{ij}$ being sensitive to system parameters and lattice geometry.
Such results are consistent with a purely symmetry-motivated approach: the coupling to the frustrated spin moments breaks time reversal and inversion symmetries, which allows spin triplet pairing correlations.

\begin{figure}
    \centering
    \includegraphics[width= 0.5\linewidth]{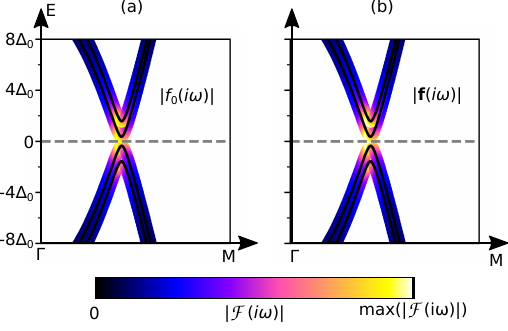}
    \caption{~BdG spectrum for the $s$-$d$ model on a triangular lattice with intra-site $s$-wave pairing.
    Excitations are colored on a continuum corresponding to the magnitude of the spin singlet pairing correlations $f_0(\mathbf{k}; i\omega)$ in (a), and the spin triplet pairing correlations $|\mathbf{f}(\mathbf{k}; i\omega)|$ in (b).
    Both spin singlet and triplet pairing correlations are shown for the intrasublattice correlation between $a$ electrons.
    Parameters for the tight-binding model are $\mu= -5 |t_0|$, $J_{sd} = |t_0|$, and $\Delta_0=0.1 |t_0|$.
    The parameters for the local spin moments are given by $\phi_0 = 0$, $\nu = 1$, and $\theta_0 = \pi/50$.
    }
    \label{fig:pairing_correlations}
\end{figure}

We demonstrate the spin triplet pairing correlations by solving exactly for the anomalous Green's functions for the lattice model of the BdG Hamiltonian in Eq.~\eqref{AppendixEq:BdG_Ham}.
As an illustrative example, we consider the model on the triangular lattice with the simplest isotropic $s$-wave-like pairing gap function.
We consider the three spins comprising the three sublattice system being in a $120^\circ$ ordered state with out-of-plane canting.
The three local spins are given by
\begin{equation}
    \begin{aligned}
    \hat{\mathbf{s}}_a &= \left(\cos \theta_0 \cos  \varphi_0 , \cos \theta_0 \sin \varphi_0, \sin \theta_0\right)^\mathrm{T},
    \\
    \hat{\mathbf{s}}_b &= \left(\cos \theta_0 \cos  \left(\varphi_0 + \nu \frac{2\pi}{3} \right) , \cos \theta_0 \sin \left(\varphi_0 + \nu \frac{2\pi}{3} \right), \sin \theta_0\right)^\mathrm{T},
    \\
    \hat{\mathbf{s}}_c &= \left(\cos \theta_0 \cos  \left(\varphi_0 - \nu \frac{2\pi}{3} \right) , \cos \theta_0 \sin \left(\varphi_0 - \nu \frac{2\pi}{3} \right), \sin \theta_0\right)^\mathrm{T},
    \end{aligned}
    \label{appendixEq:three_spins}
\end{equation}
in which $\varphi_0$ is a constant, $\theta_0$ describes the out-of-plane canting, and $\nu = \pm 1$ is the chirality for the local spins.
As an example, we numerically solve for the intrasublattice correlations for electrons at sublattice $a$, but similar results hold for the intersublattice pairing correlations.
In Figs.~\ref{fig:pairing_correlations}(a) and (b), we show the magnitude of the singlet and triplet pairing correlation magnitudes respectively for the model on a triangular lattice.
As seen in Fig.~\ref{fig:pairing_correlations}, the magnitude of the spin singlet and spin triplet pairing correlations are approximately equal.
While the relative magnitude and direction of $\mathbf{f}_{ij}(\mathbf{k})$ is largely dependent on the system parameters, including the underlying local exchange field and relative strength of $J_{sd}$, the general features---namely the coexistence of spin singlet and spin triplet superconducting pair correlations---are robust over a wide range of parameters.
{\color{black} A more detailed analysis of the pairing correlations can be found in Ref.~\citenum{Frazier2025b}.}

%% file: supplemental/s4.effective_tunneling.tex
\section{Effective tunneling}
\label{SM:tunneling}

In this section, we derive the effective tunneling of itinerant electrons in the presence of a local exchange field comprised of frustrated spin moments.
We focus on the effective tunneling between neighbouring sites $i$ and $j$, as shown schematically in Fig.~\ref{fig:tunneling} in the main text.


Consider the Hamiltonian $\mathcal{H} = \mathcal{H}_0 + \Sigma$.
Here, $\mathcal{H}_0$ denotes the Hamiltonian of the unperturbed system, and $\Sigma$ the perturbation, corresponding to the nearest neighbour hopping, $\mathcal{H}_\mathrm{kin}$, and $s$-$d$ exchange, $\mathcal{H}_{sd}$, respectively in the context of this work.
The corresponding Green's function in frequency space is given by
\begin{math}
    \mathcal{G}(i\omega) = (i\omega - \mathcal{H})^{-1}.
\end{math}
One can perform a Dyson series expansion in $\Sigma$ to construct an effective Hamiltonian.
Let $\mathcal{G}_0(i\omega) = (i\omega - \mathcal{H}_0)^{-1}$ be the Green's function of the unperturbed Hamiltonian.
From Dyson's equation, it follows that~\cite{Abrikosov1963, Balatsky2006}
\begin{equation}
    \mathcal{G}(i\omega) = (i\omega - \mathcal{H}_0 - \Sigma)^{-1}
    =
    \mathcal{G}_0(i\omega) + \mathcal{G}_0(i\omega) \Sigma  \mathcal{G}(i\omega).
\end{equation}
This can be reexpressed as
\begin{equation}
    \mathcal{G}(i\omega) = \mathcal{G}_0(i\omega) + \mathcal{G}_0(i\omega) \tilde{\Sigma}(i\omega) \mathcal{G}_0(i\omega),
\end{equation}
with $\tilde{\Sigma}$ serving as the $T$-matrix, defined recursively as
\begin{equation}
    \tilde{\Sigma}(i\omega) = \Sigma + \Sigma \mathcal{G}_0(i\omega)\tilde{\Sigma}(i\omega).
\end{equation}
In this work, we treat $s$-$d$ exchange $\mathcal{H}_{sd}$ perturbatively and retain terms up to third order in $J_{sd}$ to capture any nonvanishing spin chirality of the three-sublattice system.
From the above expansion, the effective $s$-$d$ exchange to third order in the coupling is given by
\begin{equation}
    \tilde{\Sigma}_\mathrm{eff}(i\omega)
    \approx
    \mathcal{H}_{sd}
    +
    \mathcal{H}_{sd} \mathcal{G}_\mathrm{kin} (i\omega)
    \mathcal{H}_{sd}
    +
    \mathcal{H}_{sd}
    \mathcal{G}_\mathrm{kin}(i\omega)
    \mathcal{H}_{sd}
    \mathcal{G}_\mathrm{kin} (i\omega)
    \mathcal{H}_{sd}.
    \label{appendixEq:sd_T_matrix_expansion}
\end{equation}
Above, $\mathcal{G}_\mathrm{kin}(i\omega) = (i\omega - \mathcal{H}_\mathrm{kin})^{-1}$ is the free Green's function in the absence of $s$-$d$ exchange.

Now, we consider the off-diagonal matrix elements describing the effective tunneling processes between nearest neighbouring sites.
Considering the spin-independent nearest neighbour hopping and the $T$-matrix expansion of the $s$-$d$ exchange in
Eq.~\eqref{appendixEq:sd_T_matrix_expansion}, the effective tunneling in the spin-$1/2$ basis between nearest neighbouring sites $i$ and $j$ is given by
\begin{align}
    T_{ij}
    =
    t_0 \sigma^0
    +
    J_{sd}^2
    [\mathcal{G}_\mathrm{kin}(
    i\omega)]_{i j}
    \Big(
    \left(\mathbf{s}_{i} \cdot \boldsymbol{\sigma} \right)
    \left(\mathbf{s}_{j} \cdot \boldsymbol{\sigma} \right)
    \Big)
    +
    J_{sd}^3 
    \sum_{k}
    \Big (
    [\mathcal{G}_\mathrm{kin}(
    i\omega)]_{i,k}
    [\mathcal{G}_\mathrm{kin}(
    i\omega)]_{k,j}
    \left(\mathbf{s}_{i} \cdot \boldsymbol{\sigma} \right)
    \left(\mathbf{s}_{k} \cdot \boldsymbol{\sigma} \right)
    \left(\mathbf{s}_{j} \cdot \boldsymbol{\sigma} \right)
    \Big).
\end{align}
Above, $\mathbf{s}_{i}$ is the local spin moment at site $i$, and the summation is taken over sites $k$ which are nearest neighbours to both sites $i$ and $j$.
We have used the fact that $\mathcal{G}_\mathrm{kin}(i\omega)$ is spin-independent and that the $s$-$d$ exchange is on-site.
We employ the following identities of the Pauli matrices,
\begin{gather*}
    \sigma^i \sigma^j = \delta_{ij} + i \epsilon_{ijk} \sigma_{k},
    \\
    \sigma^i \sigma^j \sigma^k
    =
    \Big( \delta_{ij} \sigma^k - \delta_{ik} \sigma^j + \delta_{jk} \sigma^i \Big)
    + i \epsilon_{ijk}.
\end{gather*}
It follows that the effective tunneling is given by
\begin{equation}
    \begin{aligned}
        T_{i j}(i\omega)
        = 
        & \ t_0 
        \sigma^0
        +
        J_{sd}^2
        [\mathcal{G}_\mathrm{kin}(i\omega)]_{i j}
        \alpha_{ij}
        \sigma^0
        +
        i J_{sd}^2
        [\mathcal{G}_\mathrm{kin}(i\omega)]_{i j}
        (\boldsymbol{\beta}_{ij} \cdot
        \boldsymbol{\sigma})
        \\
        &- 
        i J_{sd}^3
        \sum_{k}
        [\mathcal{G}_\mathrm{kin}(i\omega)]_{i k}
        [\mathcal{G}_\mathrm{kin}(i\omega)]_{k j}
        \chi_{ijk}
        \sigma^0
        +
        J_{sd}^3
        \sum_{k}
        [\mathcal{G}_\mathrm{kin}(i\omega)]_{i k}
        [\mathcal{G}_\mathrm{kin}(i\omega)]_{k j}
        \left(
        \boldsymbol{\gamma}_{ijk}
        \cdot
        \boldsymbol{\sigma}
        \right).
    \end{aligned}
    \label{appendixEq:effective_tunneling_Jsd_expansion}
\end{equation}
in which the factors,
\begin{equation}
    \begin{gathered}
        \alpha_{ij} \equiv \mathbf{s}_{i} \cdot \mathbf{s}_{j};
        \hspace{2em}
        \boldsymbol{\beta}_{ij} \equiv \mathbf{s}_{i} \times \mathbf{s}_{j};
        \\
        \chi_{ijk} \equiv \mathbf{s}_{i} \cdot (\mathbf{s}_{j}\times \mathbf{s}_{k});
        \\
        \boldsymbol{\gamma}_{ijk} \equiv
        (\mathbf{s}_{i} \cdot \mathbf{s}_{k})\mathbf{s}_{j}
        -
        (\mathbf{s}_{i} \cdot \mathbf{s}_{j})\mathbf{s}_{k}
        +
        (\mathbf{s}_{j} \cdot \mathbf{s}_{k})\mathbf{s}_{i},
    \end{gathered}
\end{equation}
describe the underlying frustration of the local spin moments, as described in the main text.

Furthermore, spin-orbit coupling can naturally enter into the effective tunneling matrices~\cite{Geshkenbein1986, Millis1988, Sigrist1991}.
For example, for systems with strong-spin orbit coupling, the spin-orbit coupling would enter into the Green's functions for itinerant electrons.
This naturally gives rise to additional spin- and momentum-dependent tunneling.
In this work, we neglect the contribution from spin-orbit coupling in order to focus on the contribution from the frustrated spin textures.
Nonetheless, the qualitative features, specifically the Josephson couplings arising from the effective tunneling as discussed in Sec.~\ref{SM:josephson}, will persist.

%% file: supplemental/s5.effective_spin_exchange.tex
\section{Effective Josephson coupling in three-sublattice system}

\label{SM:effective_exchange_interaction}

Let us consider the Josephson coupling between two grains $n$ and $m$, for which the pairing correlation functions are given by $\mathcal{F}_n$ and $\mathcal{F}_m$ respectively.
We focus on the spin degrees of freedom in determining the Josephson couplings.
One can decompose the tunneling into spin-independent and spin-dependent parts, $T_{n m} = T_{nm; 0} \sigma^0 + \mathbf{T}_{nm} \cdot \boldsymbol{\sigma}$.
The total tunneling consists of a summation of the microscopic nearest-neighbour tunneling processes across the grain boundary.
From the nearest neighbour effective tunneling in Eq.~\eqref{appendixEq:effective_tunneling_Jsd_expansion}, the spin-independent and spin-dependent tunneling of itinerant electrons across the grain boundary are given by
\begin{subequations}
    \begin{gather}
        T_{nm; 0}
        =
        \underset{j \in \Sigma_m}{\sum_{i \in \Sigma_n,}}
        T_{ij; 0}
        =
        \underset{j \in \Sigma_m}{\sum_{i \in \Sigma_n,}}
        \bigg(
        t_0 
        +
        J_{sd}^2 
        [\mathcal{G}_\mathrm{kin}]_{i j} 
        \alpha_{ij}
        +
        iJ_{sd}^3 \sum_{k} [\mathcal{G}_\mathrm{kin}]_{i k} [\mathcal{G}_\mathrm{kin}]_{k j}
        \chi_{ikj}
        \bigg),
        \\
        \mathbf{T}_{nm}
        =
        \underset{j \in \Sigma_m}{\sum_{i \in \Sigma_n,}}
        \mathbf{T}_{ij}
        =
        \underset{j \in \Sigma_m}{\sum_{i \in \Sigma_n,}}
        \bigg(
        i J_{sd}^2 [\mathcal{G}_\mathrm{kin}]_{i j} 
        \boldsymbol{\beta}_{ij}
        +
        J_{sd}^3
        \sum_{k}
        [\mathcal{G}_\mathrm{kin}]_{i k} [\mathcal{G}_\mathrm{kin}]_{k j}
        \boldsymbol{\gamma}_{ijk}
        \bigg).
    \end{gather}
\end{subequations}
Here, summation of nearest neighbouring sites $i$ and $j$ is taken over the grain boundaries $\Sigma_n$ and $\Sigma_m$ of the $n^\mathrm{th}$ and $m^\mathrm{th}$ grains respectively.

As an illustrative example, let us consider the model on a triangular lattice, shown schematically in Fig.~\ref{fig:tunneling} in the main text.
To demonstrate the relative magnitudes, we take $[\mathcal{G}_\mathrm{kin}]_{i j} \approx - (1/t_0) \sigma^0$ for nearest neighbours $i$ and $j$, which is valid for states near the Fermi surface.
For the three sublattice system, nearest neighbours correspond to sublattices with spins $\mathbf{s}_{i}$ and $\mathbf{s}_{j}$, and the summation in the third order terms corresponds to the third sublattice, which we label as $\mathbf{s}_{k}$.
The effective spin-independent and spin-dependent tunneling between nearest neighbours for the triangular lattice is approximately
\begin{subequations}
    \begin{align}
        T_{nm; 0}
        &\approx
        \underset{j \in \Sigma_m}{\sum_{i \in \Sigma_n,}}
        \bigg(
        t_0  {-} \frac{J_{sd}^2}{t_0} 
        \alpha_{ij}
        -
        2i \frac{J_{sd}^3}{t_{0}^2}
        \chi_{ijk}
        \bigg),
        \\
        \mathbf{T}_{nm}
        &\approx
        \underset{j \in \Sigma_m}{\sum_{i \in \Sigma_n,}}
        \bigg(
        {-}i \frac{J_{sd}^2}{t_0} 
        \boldsymbol{\beta}_{ij}
        +
        2 \frac{J_{sd}^3}{t_0^2}
        \boldsymbol{\gamma}_{ijk}
        \bigg).
    \end{align}
    \label{appendixEq:spin_dependent+independent_tunneling}
\end{subequations}
Here, $i$, $j$, and $k$ refer to the three sites comprising the three sublattices, and we have used $\chi_{ijk} = - \chi_{ikj}$.

We now derive the Josephson couplings between spin triplet pairing correlations in Eq.~\eqref{appendixEq:Josephson_free_energy_spin_triplet},
using the form of the tunneling in Eq.~\eqref{appendixEq:spin_dependent+independent_tunneling}.
The Josephson coupling is given by summing the different tunneling contributions across the grain boundary.
From Eq.~\eqref{appendixEq:josephson_couplings},
the effective Josephson couplings for the model on the three sublattice system are given by
\begin{subequations}
\begin{align}
    J_{nm}
    &=
    w(E_n, E_m; \beta)
    \underset{j \in \Sigma_m}{\sum_{i \in \Sigma_n,}}
    \bigg\{
    \bigg( 
    t_0 {-} \frac{J_{sd}^2}{t_0} \alpha_{ij}
    -2i \frac{J_{sd}^3}{t_0^2} \chi_{ijk}
    \bigg)^2
    +
    \bigg( {-} i\frac{J_{sd}^2}{t_0} \boldsymbol{\beta}_{ij}
    +
    2
    \frac{J_{sd}^3}{t_0^2}
    \boldsymbol{\gamma}_{ijk}\bigg)^2
    \bigg\},
\\
    {\mathbf{D}}_{nm}
    &= 
    2i 
    w(E_n, E_m; \beta)
    \underset{j \in \Sigma_m}{\sum_{i \in \Sigma_n,}}
    \bigg\{
    \left( 
    t_0 {-} \frac{J_{sd}^2}{t_0} \alpha_{ij}
    -2i \frac{J_{sd}^3}{t_0^2} \chi_{ijk}
    \right)
    \left( {-} i\frac{J_{sd}^2}{t_0} \boldsymbol{\beta}_{ij}
    +
    2
    \frac{J_{sd}^3}{t_0^2}
    \boldsymbol{\gamma}_{ijk}\right)
    \bigg\},
\\
    {\Gamma}_{nm}^{ab}
    &= 
    -2
    w(E_n, E_m; \beta)
    \underset{j \in \Sigma_m}{\sum_{i \in \Sigma_n,}}
    \bigg\{
    \left( {-} i\frac{J_{sd}^2}{t_0} \beta_{ij}^a
    +
    2
    \frac{J_{sd}^3}{t_0^2}
    \gamma_{ijk}^a\right)
    \left( {-} i\frac{J_{sd}^2}{t_0} \beta_{ij}^b
    +
    2
    \frac{J_{sd}^3}{t_0^2}
    \gamma_{ijk}^b\right)
    \bigg\}.
\end{align}
\label{appendixEq:josephson_coupling_amplitudes_unexpanded}
\end{subequations}
To lowest orders in the $sd$-coupling, the effective Josephson couplings can be approximated by
\begin{subequations}
    \begin{align}
        {J}_{nm} 
        & \approx 
        - \frac{\Delta_0}{W^2}
        \underset{j \in \Sigma_m}{\sum_{i \in \Sigma_n,}}
        \bigg\{
        t_0^2  {-} 2 J_{sd}^2 \alpha_{ij} - 4i \frac{J_{sd}^3}{t_0} \chi_{ijk}
        \bigg \}
        + \mathcal{O}(J_{sd}^4),
        \\
        {\mathbf{D}}_{nm}
        &\approx
        -\frac{2\Delta_0}{W^2}
        \underset{j \in \Sigma_m}{\sum_{i \in \Sigma_n,}}
        \bigg\{
        J_{sd}^2 \boldsymbol{\beta}_{ij}
        +
        2i \frac{J_{sd}^3}{t_0} \boldsymbol{\gamma}_{ijk}
        \bigg\}
        + \mathcal{O}(J_{sd}^4),
        \\
        {\Gamma}_{nm}^{ab}
        &\approx
        - \frac{2 \Delta_0}{W^2} \underset{j \in \Sigma_m}{\sum_{i \in \Sigma_n,}}
        \frac{J_{sd}^4}{t_0^2} \beta_{ij}^a \beta_{ij}^b
         + \mathcal{O}(J_{sd}^5),
    \end{align}
\end{subequations}
up to an overall factor.
{\color{black} Above, we have used the $w(E_n, E_m; \beta) \approx -\Delta_0/W^2,$ in which $W$ is the bandwidth.
In the nonperturabative regime, all terms in Eq.~\eqref{appendixEq:josephson_coupling_amplitudes_unexpanded} must be considered, which can lead to additional corrections to the coupling amplitudes~\cite{Frazier2025b}.
}

In the limit of weak or vanishing coupling to the exchange field, the Heisenberg-like $J_{nm}$ term dominates, promoting {collinear} $d$ vectors at adjacent sites.
In this limit, minimization of the free energy leads to a homogeneous order parameter, in which any variation from site to site is penalized.
In contrast, when $J_{sd}$ is finite, it follows that  Dzyaloshinskii-Moriya-like coupling $\mathbf{D}_{nm}$ becomes significant and competes with $J_{nm}$.
This competition promotes {noncollinear} configurations of $d$ vectors at adjacent sites, leading to inhomogeneous real-space textures of the $d$ vector, such as skyrmion-like excitations.
Moreover, as the $d$ vector can vary from site to site, this can lead to nontrivial contributions to the superfluid velocity depending on the underlying exchange field.
In the following, we analyze three representative cases illustrating how different underlying spin configurations affect the Josephson coupling in the presence of finite $J_{sd}$.

\subsubsection*{Case 1: Collinear Spins}

For the first case, suppose all local spins are in a ferromagnetic state, \textit{i.e.} the spins are not frustrated and fully collinear.
In this case, $\alpha_{ij} = \mathbf{s}_i \cdot \mathbf{s}_j$ is maximized, while the measure of noncollinearity $\boldsymbol{\beta}_{ij} = \mathbf{s} \times \mathbf{s}_j$ and spin chirality $\chi_{ijk} = \mathbf{s}_i \cdot (\mathbf{s}_j \times \mathbf{s}_k)$ are zero.
It is to be noted that the third order term, $\boldsymbol{\gamma}_{ijk}$ is nonvanishing, even for collinear spins.
Because the spins are collinear, the DM-like coupling $\mathbf{D}_{nm}$ is the same for each grain and points in the direction of local ferromagnetic order.
However, because the DM-like term originates from a higher order scattering process, it plays a less significant role in the Josephson free energy.
Consequently, for collinear spin texture with weak $s$-$d$ exchange, the Heisenberg-like Josephson coupling $J_{nm}$ dominates.

\subsubsection*{Case 2: Coplanar Spins}

Next, we consider when the localized spins of the $d$ electrons are in the classical groundstate for the three-sublattice system, a $120^\circ$ ordered coplanar antiferromagnet.
This corresponds to canting angle $\theta_0 = 0$ in Eq.~\eqref{appendixEq:three_spins}.
For coplanar spins, the scalar spin chirality vanishes, $\chi_{ijk} = 0$, while $\alpha_{ij}$ and $\beta_{ij}$ are both finite.
To lowest order in the $J_{sd}$ coupling, the nonvanishing terms in Josephson free energy in Eq.~\eqref{appendixEq:Josephson_free_energy_spin_triplet} are given by 
\begin{equation}
    {J}_{nm} \approx 
    -\frac{\Delta_0}{W^2}
    \underset{j \in \Sigma_m}{\sum_{i \in \Sigma_n,}}
    (
    t_0^2 {-} J_{sd}^2
    )
    , \hspace{3em}
    {\mathbf{D}}_{nm} \approx
    {-} \frac{\Delta_0}{W^2}
    \underset{j \in \Sigma_m}{\sum_{i \in \Sigma_n,}}
    (\sqrt{3} J_{sd}^2 \nu
    \epsilon_{ij}
    \hat{z}
    ).
\end{equation}
Above, we have used $\alpha_{ij} = \cos 2\pi/3 = -1/2$ and $\boldsymbol{\beta}_{ij} = (\sqrt{3}/2) \epsilon_{ij} \hat{z}$ for the $120^{\circ}$ ordered classical spin moments, with $\epsilon_{ij}$ being the antisymmetric tensor.
For nonvanishing $J_{sd}$, the superfluid stiffness $J_{nm}$ competes with the DM-like Josephson coupling $D_{nm}$ favoring noncollinear $d$ vectors at adjacent grains, leading to a spatially inhomogeneous superconducting pairing order.

\subsubsection*{Case 3: Nonvanishing spin chirality}

Lastly, we consider the case where the three $120^\circ$ ordered spins in the ground state are canted out-of-plane by angle $\theta_\mathrm{0}$, as given by Eq.~\eqref{appendixEq:three_spins}.
The spin chirality is nonvanishing for finite $\theta_0$,
\begin{equation}
    \chi_{abc} = \hat{\mathbf{s}}_a \cdot (\hat{\mathbf{s}}_b \times \hat{\mathbf{s}}_c)
    =
    \nu \frac{3 \sqrt{3}}{2} \cos^2 \theta_0 \sin \theta_0,
\end{equation}
and is dependent on the sign of the canting and the chirality, $\nu$.
Similarly, the DM-vector is tilted away from the $z$ axis and is dependent on the sublattices at the grain boundary,
\begin{equation}
    \begin{aligned}
        \boldsymbol{\beta}_{ab} 
        &= \hat{\mathbf{s}}_a \times \hat{\mathbf{s}}_b 
        = 
        \frac{1}{2}
        \left(
        \begin{array}{c}
             \sin(2\theta_0) \Big( \frac{3}{2} \sin \varphi_0 -  \frac{\sqrt{3}}{2} \nu \cos \varphi_0 \Big)
             \\
             - \sin(2\theta_0) \Big(\frac{3}{2} \cos \varphi_0 +  \frac{\sqrt{3}}{2} \nu \sin \varphi_0 \Big)
             \\
             \nu \sqrt{3} \cos^2 \theta_0
        \end{array}
        \right)
        \\
        \boldsymbol{\beta}_{bc} 
        &= \hat{\mathbf{s}}_b \times \hat{\mathbf{s}}_c 
        = 
        \frac{\nu \sqrt{3}}{2}
        \left(
        \begin{array}{c}
             \sin(2\theta_0) \cos \phi_0
             \\
             \sin(2\theta_0) \sin \phi_0
             \\
             \cos^2 \theta_0
        \end{array}
        \right)
        \\
        \boldsymbol{\beta}_{ca} 
        &= \hat{\mathbf{s}}_c \times \hat{\mathbf{s}}_a
        = 
        \frac{1}{2}
        \left(
        \begin{array}{c}
             - \sin(2\theta_0) \Big( \frac{3}{2} \sin \varphi_0 +  \frac{\sqrt{3}}{2} \nu \cos \varphi_0 \Big)
             \\
             \sin(2\theta_0) \Big( 
             \frac{3}{2} \cos \varphi_0 - \frac{\sqrt{3}}{2} \nu \sin \varphi_0
             \Big)
             \\
             \nu \sqrt{3} \cos^2 \theta_0
        \end{array}
        \right).
    \end{aligned}
\end{equation}
For nonvanishing spin chirality, the DM-vector changes sign in real space depending on the interlayer sublattice tunneling to lowest order.
To minimize the free energy functional, the $d$ vector will gain an out-of-plane component which will vary from grain to grain.

%% file: supplemental/s6.vortices.tex
\section{Contribution to superfluid velocity from \texorpdfstring{$d$}{d} vector texture}

\label{SM:superfluid_velocity_d_vector}

In this section, we derive the contribution to the superfluid velocity and its circulation from the spatial texture of the $d$ vector field.
To highlight the role of the $d$ vector, we assume that the system does not conserve total angular momentum, so that the orbital and spin degrees of freedom may be decoupled.
Consider the spin triplet pairing order described by 
\begin{math}
    \Delta_{ij} (\mathbf{r}) = |\Delta_{ij}(\mathbf{r})| \hat{\Delta}_{ij}(\mathbf{r}),
\end{math}
in which
\begin{equation}
    \hat{\Delta}_{ij} = e^{i\phi(\mathbf{r})} (\hat{m}_i(\mathbf{r}) + i \hat{n}_i(\mathbf{r})) \hat{d}_j (\mathbf{r}) \sigma^j (i\sigma^y).
\end{equation}
Above, $\hat{\mathbf{m}}(\mathbf{r})$ and $\hat{\mathbf{n}}(\mathbf{r})$ are orthogonal vectors that describe the orbital degree of freedom of the Cooper pair, while $\hat{\mathbf{d}}(\mathbf{r})$ is the $d$ vector.
In the following, we assume that the orbital angular momentum is not conserved and does not contribute to the superfluid velocity.
The pairing is defined as unitary if it satisfies $\hat{\Delta} \hat{\Delta}^\dagger = \mathbbm{1}$, or equivalently $\hat{\mathbf{d}}(\mathbf{r}) \times \hat{\mathbf{d}}^*(\mathbf{r}) = 0.$
Generally, one can decompose the $d$ vector into  purely unitary and purely nonunitary components,
\begin{equation}
    \hat{\mathbf{d}}(\mathbf{r}) = \mathbf{d}_U(\mathbf{r}) + \mathbf{d}_N(\mathbf{r}).
\end{equation}
Here, $\hat{\mathbf{d}}_U(\mathbf{r})$ is real valued and satisfies $\hat{\mathbf{d}}_U(\mathbf{r}) \cdot \hat{\mathbf{d}}_U(\mathbf{r}) = 1$, and $\hat{\mathbf{d}}_N(\mathbf{r})$ is complex-valued and satisfies 
$|\hat{\mathbf{d}}_N \times \hat{\mathbf{d}}_N^*| = 1$.
This decomposition into a unitary and nonunitary pairing order parameter is in general not unique.
As an example, the $d$ vector can be decomposed as
\begin{equation}
    {\mathbf{d}}_U(\mathbf{r}) = \frac{\mathrm{Re}[\hat{\mathbf{d}}(\mathbf{r})]\cdot \mathrm{Im}[\hat{\mathbf{d}}(\mathbf{r})]}{|\mathrm{Im}[\hat{\mathbf{d}}(\mathbf{r})]|^2}
    \mathrm{Im}[\hat{\mathbf{d}}(\mathbf{r})]
    +
    i\mathrm{Im}[\hat{\mathbf{d}}(\mathbf{r})]
    ; \hspace{2em}
    {\mathbf{d}}_N(\mathbf{r}) = \hat{\mathbf{d}}(\mathbf{r}) - {\mathbf{d}}_U(\mathbf{r}).
\end{equation}
One can always choose a convention such that $\mathbf{d}_U(\mathbf{r})$ is real-valued by absorbing its complex phase into $\phi(\mathbf{r})$.
Now, let us decompose $\hat{\mathbf{d}}_N$ into its real and imaginary parts,
\begin{equation}
    \hat{\mathbf{d}}_N(\mathbf{r}) = \mathbf{a}(\mathbf{r}) +  i \mathbf{b}(\mathbf{r}),
\end{equation}
in which ${\mathbf{a}}(\mathbf{r})$ and ${\mathbf{b}}(\mathbf{r})$ are real-valued.
Because $\hat{\mathbf{d}}_N$ is purely nonunitary, it follows that $\hat{\mathbf{d}}_N(\mathbf{r}) \cdot \hat{\mathbf{d}}_N(\mathbf{r}) = 0$.
As such, $\mathbf{a}$ and $\mathbf{b}$ are orthogonal and of equal magnitude, satisfying $|{\mathbf{a}}(\mathbf{r}) \times {\mathbf{b}}(\mathbf{r})| = 1/2$, with $|\mathbf{a}| = |\mathbf{b}| = 1/\sqrt{2}$.
One can form a local orthonormal triad $\{\hat{\mathbf{a}}(\mathbf{r}), \hat{\mathbf{b}}(\mathbf{r}), \hat{\mathbf{S}}(\mathbf{r})\}$, in which $\hat{\mathbf{S}}(\mathbf{r}) = i \hat{\mathbf{d}}_N(\mathbf{r}) \times \hat{\mathbf{d}}^*_N(\mathbf{r}) = \hat{\mathbf{a}}(\mathbf{r}) \times \hat{\mathbf{b}}(\mathbf{r})$ is the unit vector pointing in the local direction of the spin of the Cooper pair.

The superfluid velocity is generally given by
\begin{equation}
    \mathbf{v}_s = - \frac{i}{2} \frac{\hbar}{m^*}
    ( \hat{\Delta}^\dagger_{j}(\mathbf{r}) \boldsymbol{\nabla} \hat{\Delta}_{j}(\mathbf{r}) - \mathrm{h.c.}),
\end{equation}
with the summation implied over repeated index $j=1,2,3$ and $m^*$ being the mass of the Cooper pair.
For spin triplet pairing order parameters, the superfluid velocity can be expressed as
\begin{equation}
    v_{s,j} = \frac{\hbar}{m^*}
    \Big(
    \partial_j \phi(\mathbf{r})
    -i \hat{\mathbf{d}}^*(\mathbf{r}) \cdot \partial_j \hat{\mathbf{d}}(\mathbf{r}) 
    \Big),
    \label{appendixEq:superfluid_velocity_d_vector}
\end{equation}
which includes a contribution from the spatial variation of the $\mathrm{U}(1)$ phase and also from the spatial texture of the $d$ vector.
The latter contribution can be viewed as a $\mathrm{U}(1)$ gauge field $A_j(\mathbf{r}) = i \hat{\mathbf{d}}^* \cdot \partial_j \hat{\mathbf{d}}$ and is nonvanishing for nonunitary pairing orders.

Let us consider the case in which the $\mathrm{U}(1)$ phase $\phi(\mathbf{r})$ has no singularities.
In this case, the curl of the superfluid velocity is solely determined by the spatial texture of the $d$ vector,
\begin{align}
    (\boldsymbol{\nabla} \times \mathbf{v}_s)_i
    &=
    -i \frac{\hbar}{m^*} \epsilon_{ijk} \partial_j\hat{\mathbf{d}}^*(\mathbf{r}) \cdot \partial_k \hat{\mathbf{d}}(\mathbf{r})
    =
    -i \frac{\hbar}{m^*} \epsilon_{ijk} \partial_j A_k.
    \label{appendixEq:curl_of_velocity}
\end{align}
One can express the right hand side of Eq.~\eqref{appendixEq:curl_of_velocity} in terms of the gauge-invariant quantity $\hat{\mathbf{S}}(\mathbf{r})$.
To see, consider a purely nonunitary pairing order parameter, with $\hat{\mathbf{d}}(\mathbf{r}) = \hat{\mathbf{d}}_N(\mathbf{r})$.
\begin{equation}
    \hat{\mathbf{d}}^*_N(\mathbf{r})
    \cdot \partial_j
    \hat{\mathbf{d}}_N(\mathbf{r})
    =
    2i \mathbf{a}(\mathbf{r}) \cdot \partial_j \mathbf{b}(\mathbf{r})
    =
    i \hat{\mathbf{a}}(\mathbf{r}) \cdot \partial_j \hat{\mathbf{b}}(\mathbf{r}).
\end{equation}
As such, the superfluid velocity in Eq.~\eqref{appendixEq:superfluid_velocity_d_vector} can be expressed as
\begin{equation}
    v_{s,j} = \frac{\hbar}{m^*}
    \Big(
    \partial_j \phi(\mathbf{r})
    + \hat{\mathbf{a}}(\mathbf{r})\cdot \partial_j \hat{\mathbf{b}}(\mathbf{r})
    \Big).
    \label{appendixEq:superfluid_velocity_nonunitary_d_vector}
\end{equation}
To compare, for superfluid $^3$He-$A$, whose pairing order is given by~\cite{Anderson1973, Vollhardt2013}
\begin{math}
    \Delta_{\mu j}(\mathbf{r})=|\Delta_0(\mathbf{r})| \hat{d}_\mu (\mathbf{r}) (\hat{m}_j (\mathbf{r}) +i \hat{n}_j (\mathbf{r})),
\end{math}
with $d$ vector being real-valued and $\hat{\mathbf{m}}(\mathbf{r}) \times \hat{\mathbf{n}}(\mathbf{r}) = \hat{\mathbf{l}}(\mathbf{r})$
pointing in the local direction of the nonvanishing angular momentum of the Cooper pair.
The superfluid velocity of $^3$He-$A$ is given by
\begin{equation}
    v_{s, j} =
    \frac{\hbar}{m^*}
    \Big(
    \partial_j \phi(\mathbf{r})
    +
    \hat{\mathbf{m}}(\mathbf{r}) \cdot \partial_j\hat{\mathbf{n}}(\mathbf{r})
    \Big).
\end{equation}
This is analogous to the superfluid velocity for the nonunitary spin triplet pairing order in Eq.~\eqref{appendixEq:superfluid_velocity_nonunitary_d_vector}, now with the orthonormal triad $\{\hat{\mathbf{a}}, \hat{\mathbf{b}}, \hat{\mathbf{S}}\}$ playing the role of $\{\hat{\mathbf{m}}, \hat{\mathbf{n}}, \hat{\mathbf{l}}\}$ respectively.
Following the standard derivation of the Mermin-Ho relation~\cite{Mermin1976, Kamien2002},
the curl of superfluid velocity is given by~\cite{Salomaa1987, Volovik2009}
\begin{equation}
    (\boldsymbol{\nabla} \times \mathbf{v}_s)_i = \frac{1}{2} \frac{\hbar}{m^*}
    \epsilon_{abc} \epsilon_{ijk} \hat{S}_a \partial_j \hat{S}_b \partial_k \hat{S}_c,
\end{equation}
for nonunitary spin triplet superconductors.
Here, the local spin of the Cooper pair $\hat{\mathbf{S}}(\mathbf{r})$ plays the role of the $l$ vector in the conventional Mermin-Ho relation.


%% file: supplemental/s7.majorana.tex
\section{Contribution from Majorana bound states}

{\color{black}
In this section, we discuss possible Majorana bound states and their contribution to the Josephson effects.
We consider spin triplet pairing order with odd partial wave symmetry, which is known to be able to host zero-energy Majorana bound states~\cite{Read2000, Kitaev2001, Alicea2012}.
Below, we consider the role of such states in a spin triplet superconducting system for a varying $d$ vector texture.

\subsection{Majorana edge states for spatially varying \texorpdfstring{$d$}{d} vector texture}

As discussed in the main text, we consider a smoothly varying $d$ vector texture arising from anisotropic Josephson couplings in the presence of a frustrated spin texture.
For a smoothly varying $d$ vector texture, which does not change sign, the Majorana states are confined to the edges of the sample, provided the bulk pairing gap does not close.

To demonstrate, we consider a one-dimensional spinful Kitaev chain~\cite{Kitaev2001}.
The mean-field Hamiltonian is given by
\begin{equation}
    \begin{aligned}
        H 
        &= -\mu \sum_{i,\alpha} c^\dagger_{i, \alpha} c_{i, \alpha}
        +
        t_0 \sum_{i, \alpha}
        \Big(c^\dagger_{i+1, \alpha} c_{i, \alpha} + c^\dagger_{i, \alpha} c_{i+1, \alpha}\Big)
        \\
        &\hspace{1em}
        +
        \sum_{i, \alpha, \beta}
        \left(
        \Delta_0
        [(\hat{\mathbf{d}}(x_i) \cdot \boldsymbol{\sigma})i\sigma_y]_{\alpha \beta}
        \Big(
        c^\dagger_{i+1, \alpha} c^\dagger_{i, \beta}
        -
        c^\dagger_{i, \alpha} c^\dagger_{i+1, \beta}\Big)
        +
        \mathrm{h.c.}
        \right)
    \end{aligned}
    \label{appendixEq:kitaev_and_sd}
\end{equation}
in which $t_0$ is the hopping amplitude,
$\Delta_0 = |\Delta_0| e^{i \vartheta}$ with $\vartheta$ being the overall pairing phase, and
$\hat{\mathbf{d}}(x_i)$ describes the $d$ vector at the bond between sites $x_i$ and $x_{i+1}$.
The BdG Hamiltonian kernel in Nambu basis is given by
\begin{equation}
    \mathcal{H}_\mathrm{BdG} = 
    \left(
    \begin{array}{cc}
        \mathcal{H} & \Delta
        \\
        \Delta^\dagger & -\mathcal{H}^\mathrm{T}
    \end{array}
    \right),
\end{equation}
in which the normal-metal Hamiltonian, $\mathcal{H}$, includes nearest-neighbour hopping,
and $\Delta$ is a pairing matrix in spin-space which has $p_x$-wave orbital pairing symmetry.
The nonvanishing matrix elements of $\mathcal{H}$ are given by
\begin{equation}
    \mathcal{H}_{i, i+1} = \mathcal{H}_{i, i-1} = t_0  \sigma_0
    ;
    \hspace{2em}
    \mathcal{H}_{i,i} = -\mu \sigma_0 
\end{equation}
and the nonvanishing matrix elements of the pairing gap function are given by
\begin{equation}
    \Delta_{i, i + 1} = - \Delta_{i+1, i}
    =
    \Delta_0 (\hat{\mathbf{d}}(x_i) \cdot \boldsymbol{\sigma})i\sigma_y.
\end{equation}

We first discuss the case for a uniform $d$ vector texture and bulk-edge correspondence, and later discuss the effects of a varying $d$ vector texture.
Without loss of generality, suppose the $d$ vector has the form $\hat{\mathbf{d}}(x_i) = \hat{\mathbf{x}}$, corresponding to equal-spin pairing $|\!\downarrow \downarrow \rangle - |\!\uparrow \uparrow\rangle$, up to an overall phase convention.
This is equivalent to two uncoupled, spin-polarized Kitaev chains.
We define Majorana operators~\cite{Kitaev2001}
\begin{equation}
    \begin{gathered}
        \gamma_{2i-1, \alpha} = 
        {e^{-i (1-\alpha) \pi/4}}
        e^{-i\vartheta/2} \tilde{c}_{i, \alpha} + 
        {e^{i (1-\alpha) \pi/4}}
        e^{i\vartheta/2} \tilde{c}_{i, \alpha}^\dagger
        \\
        \gamma_{2i, \alpha} = 
        \frac{1}{i}
        \left(
        {e^{-i (1-\alpha) \pi/4}}
        e^{-i\vartheta/2} \tilde{c}_{i, \alpha} - 
        {e^{i (1-\alpha) \pi/4}}
        e^{i\vartheta/2} \tilde{c}_{i, \alpha}^\dagger)
        \right)
    \end{gathered}
    \label{appendixEq:majorana_operators}
\end{equation}
which satisfy the real Clifford algebra $\{\gamma_{i, \alpha}, \gamma_{j, \beta}\} = 2\delta_{ij} \delta_{\alpha \beta}$.
Here, $\alpha, \beta = \pm 1$ denote spin ($\uparrow, \downarrow$) and $i,j$ take values from $1$ to $N$, with $N$ being the number of sites in the chain.
We have introduced operators ${c}_{i, \alpha} = e^{-i \vartheta/2} \tilde{c}_{i, \alpha}$ which explicitly include the superconducting $\mathrm{U}(1)$ phase $\vartheta$, and we have also accounted for the relative phase difference of the $|\!\uparrow\uparrow\rangle$ and $|\!\downarrow\downarrow\rangle$ pairing channels.
The transformation and its inverse are given by
\begin{subequations}
    \begin{align}
        \left(
        \begin{array}{c}
             \gamma_{2i-1, \alpha}
             \\
             \gamma_{2i, \alpha}
        \end{array}
        \right)
        &=
        \left(
        \begin{array}{cc}
             {e^{-i (1-\alpha) \pi/4}}
             e^{-i \vartheta/2} 
             & 
             {e^{i (1-\alpha) \pi/4}}
             e^{i \vartheta/2}
             \\
             -i 
             {e^{-i (1-\alpha) \pi/4}}
             e^{-i \vartheta/2 }
             &
             i 
             {e^{i (1-\alpha) \pi/4}}
             e^{i \vartheta/2 }
        \end{array}
        \right)
        \left(
        \begin{array}{c}
             \tilde{c}_{i, \alpha}
             \\
             \tilde{c}^\dagger_{i, \alpha}
        \end{array}
        \right)
        \\
        \left(
        \begin{array}{c}
             \tilde{c}_{i, \alpha}
             \\
             \tilde{c}^\dagger_{i, \alpha}
        \end{array}
        \right)
        &=
        \frac{1}{2}
        \left(
        \begin{array}{cc}
             {e^{i (1-\alpha) \pi/4}}
             e^{i \vartheta/2} 
             & 
             i 
             {e^{i (1-\alpha) \pi/4}}
             e^{i \vartheta/2}
             \\
             {e^{-i (1-\alpha) \pi/4}}
             e^{-i \vartheta/2 }
             &
             -i 
             {e^{-i (1-\alpha) \pi/4}}
             e^{-i \vartheta/2}
        \end{array}
        \right)
        \left(
        \begin{array}{c}
             \gamma_{2i-1, \alpha}
             \\
             \gamma_{2i, \alpha}
        \end{array}
        \right),
    \end{align}    
\end{subequations}
which yields
\begin{subequations}
    \begin{align}
        \tilde{c}_{i, \uparrow} 
        &=
        \frac{1}{2} e^{i \vartheta /2}
        \Big(
        \gamma_{2i - 1, \uparrow}
        +
        i
        \gamma_{2i, \uparrow}
        \Big)
        \\
        \tilde{c}_{i, \downarrow} 
        &=
        \frac{i}{2} e^{i \vartheta /2}
        \Big(
        \gamma_{2i - 1, \downarrow}
        +
        i
        \gamma_{2i, \downarrow}
        \Big).
    \end{align}
    \label{appendixEq:c_in_terms_of_gamma.d_along_x}
\end{subequations}
Written in terms of Majorana operators, the Hamiltonian in Eq.~\eqref{appendixEq:kitaev_and_sd} for $\hat{\mathbf{d}}(x_i) = \hat{\mathbf{x}}$
can be expressed as
\begin{equation}
    \begin{aligned}
        H 
        &= - \frac{i \mu}{2} \sum_{i,\alpha} \gamma_{2i-1, \alpha} \gamma_{2i, \alpha}
        - \sum_i \frac{\mu}{2}
        +
        \sum_{i, \alpha}
        \frac{i (t_0 + 
        \Delta_0)}{2}
        \gamma_{2i+1, \alpha} \gamma_{2i, \alpha}
        -
        \sum_{i, \alpha}
        \frac{i (t_0 - 
        \Delta_0)}{2}
        \gamma_{2i+2, \alpha} \gamma_{2i-1, \alpha}.
    \end{aligned}
\end{equation}
Above, we have used the fact that $[i \sigma_x \sigma_y]_{\alpha \alpha} = -[\sigma_z]_{\alpha \alpha} = \alpha$ for spin-up ($\alpha = 1$) and spin-down ($\alpha= -1$), as well as the following relations,
\begin{subequations}
    \begin{align}
        c^\dagger_{i \alpha} c_{j \beta}
        &=
        \frac{1}{4}
        \Big(
        \gamma_{2i-1, \alpha} \gamma_{2j-1, \beta}
        +
        \gamma_{2i, \alpha} \gamma_{2j, \beta}
        +
        i \gamma_{2i-1, \alpha} \gamma_{2j, \beta}
        -
        i \gamma_{2i, \alpha} \gamma_{2j-1, \beta}
        \Big)
        \\
        c^\dagger_{j+1, \alpha} c_{j, \alpha}
        &=
        \frac{1}{4}
        \left(
        \gamma_{2j+1, \alpha} \gamma_{2j-1, \alpha}
        +
        \gamma_{2j+2, \alpha} \gamma_{2j, \alpha}
        +
        i\gamma_{2j+1, \alpha} \gamma_{2j, \alpha} 
        - i \gamma_{2j+2, \alpha} \gamma_{2j-1, \alpha}
        \right)      
        \\
        c^\dagger_{j+1, \alpha} c^\dagger_{j, \alpha}
        -
        c_{j+1, \alpha} c_{j, \alpha}
        &=
        \frac{\alpha}{2i}
        \Big(
        \gamma_{2j+1, \alpha} \gamma_{2j, \alpha} + \gamma_{2j+2, \alpha} \gamma_{2j-1, \alpha}
        \Big).
    \end{align}
\end{subequations}
When the $d$ vector texture is uniform, $\hat{\mathbf{d}}(x_i) = \hat{\mathbf{d}}_0$,
the system is analogous to the standard Kitaev chain and features zero-energy Majorana states localized at the edges of the chain, as shown in Fig.~\ref{fig:majorana_states}(a).
There are in total four zero-energy states, with the four-fold degeneracy resulting from the spin degree of freedom, corresponding to two decoupled copies of the Kitaev chain.

For unitary pairing with $d$ vector oriented along a different direction, a unitary transformation of Eq.~\eqref{appendixEq:majorana_operators} can be used to determine the Majorana states.
In general, a real $d$ vector 
oriented along $\hat{\mathbf{n}} = (\sin \tilde{\theta} \cos \tilde{\phi}, \sin \tilde{\theta} \sin \tilde{\phi}, - \cos \tilde{\theta})^\mathrm{T}$ 
can be expressed by rotation from a $d$ vector oriented along the $x$-direction via the following transformation,
\begin{equation}
    \Delta (x_i)=
    \Delta_0
    \Big(
    \hat{\mathbf{d}}(x_i) \cdot \boldsymbol{\sigma}
    \Big)
    i \sigma_y
    =
    \Delta_0
    \Big(
    U^\dagger(x_i)
    \sigma_x
    U(x_i)
    \Big)
    i\sigma_y
    =
    \Delta_0
    U^\dagger (x_i)
    \Big(
    i \sigma_x\sigma_y
    \Big) U^*(x_i),
    \label{appendixEq:rotated_d_vector}
\end{equation}
in which
\begin{align}
    U(x_i) 
    &\nonumber
    \equiv U_z(\tilde{\phi}(x_i))
    U_y(\frac{\pi}{2}-\tilde{\theta}(x_i))
    \\
    &= \nonumber
    \left(
    \cos \frac{\tilde{\phi}(x_i)}{2} \sigma_0
    -
    i \sin \frac{\tilde{\phi}(x_i)}{2} \sigma_z
    \right)
    \left(
    \cos \left(- \frac{\pi}{4} + \frac{\tilde{\theta}(x_i)}{2}\right)
    \sigma_0
    +
    i \sin \left(- \frac{\pi}{4} + \frac{\tilde{\theta}(x_i)}{2}\right)
    \sigma_y
    \right)
    \\
    &=
    \left(
    \begin{array}{cc}
        \cos
        \left(
        - \frac{\pi}{4} + \frac{\tilde{\theta}(x_i)}{2}
        \right)
        e^{-i \tilde{\phi}(x_i)/2}
        &
        \sin
        \left(
        - \frac{\pi}{4} + \frac{\tilde{\theta}(x_i)}{2}
        \right)
        e^{-i \tilde{\phi}(x_i)/2}
        \\
        -\sin
        \left(
        - \frac{\pi}{4} + \frac{\tilde{\theta}(x_i)}{2}
        \right)
        e^{i \tilde{\phi}(x_i)/2}
        &
        \cos
        \left(
        - \frac{\pi}{4} + \frac{\tilde{\theta}(x_i)}{2}
        \right)
        e^{i \tilde{\phi}(x_i)/2}
    \end{array}
    \right)
    \label{appendixEq:rotation}
\end{align}
is a rotation about fixed axes $\hat{\mathbf{y}}$ and $\hat{\mathbf{z}}$.
Here, we have used the identity 
$U \sigma_y = \sigma_y U^*$, 
for $\mathrm{SU}(2)$ matrix $U = u_0 \sigma_0 + i \mathbf{u}\cdot \boldsymbol{\sigma}$ with $u_0$ and $\mathbf{u}$ being real-valued.

The $p_x$-wave pairing for general unitary pairing is given by
\begin{equation}
    \left(
    \begin{array}{cc}
         c_{i, \uparrow}^\dagger
         &
         c_{i, \downarrow}^\dagger
    \end{array}
    \right)
    \Delta(x_i)
    \left(
    \begin{array}{c}
         c_{i+1, \uparrow}^\dagger
         \\
         c_{i+1, \downarrow}^\dagger
    \end{array}
    \right)
    =
    \Delta_0
    e^{i\theta}
    \left(
    \begin{array}{cc}
         c_{i, \uparrow}^\dagger
         &
         c_{i, \downarrow}^\dagger
    \end{array}
    \right)
    U^\dagger(x_i)
    (\sigma_x i \sigma_y)
    U^*(x_i)
    \left(
    \begin{array}{c}
         c_{i+1, \uparrow}^\dagger
         \\
         c_{i+1, \downarrow}^\dagger
    \end{array}
    \right).
\end{equation}
Consider the pairing given by
\begin{math}
    \left(
    \begin{array}{cc}
         c_{i, \uparrow}'^\dagger
         &
         c_{i, \downarrow}'^\dagger
    \end{array}
    \right)
    (\Delta_0 \sigma_x i \sigma_y)
    \left(
    \begin{array}{c}
         c_{i+1, \uparrow}'^\dagger
         \\
         c_{i+1, \downarrow}'^\dagger
    \end{array}
    \right),
\end{math}
which corresponds to $d$ vector oriented along the $x$-direction.
From Eq.~\eqref{appendixEq:c_in_terms_of_gamma.d_along_x},
it follows that the operators are related to the rotated basis by the transformation
\begin{align}
    \left(
    \begin{array}{c}
         \tilde{c}_{i, \uparrow}^\dagger
         \\
         \tilde{c}_{i, \downarrow}^\dagger
    \end{array}
    \right)
    &= \nonumber
    [U^*(x_i)]^{-1}
    \left(
    \begin{array}{c}
         \tilde{c}_{i, \uparrow}'^\dagger
         \\
         \tilde{c}_{i, \downarrow}'^\dagger
    \end{array}
    \right)
    \\
    &= \nonumber
    \left(
    \begin{array}{cc}
        \cos
        \left(
        - \frac{\pi}{4} + \frac{\tilde{\theta}(x_i)}{2}
        \right)
        e^{-i \tilde{\phi}(x_i)/2}
        &
        -\sin
        \left(
        - \frac{\pi}{4} + \frac{\tilde{\theta}(x_i)}{2}
        \right)
        e^{i \tilde{\phi}(x_i)/2}
        \\
        \sin
        \left(
        - \frac{\pi}{4} + \frac{\tilde{\theta}(x_i)}{2}
        \right)
        e^{-i \tilde{\phi}(x_i)/2}
        &
        \cos
        \left(
        - \frac{\pi}{4} + \frac{\tilde{\theta}(x_i)}{2}
        \right)
        e^{i \tilde{\phi}(x_i)/2}
    \end{array}
    \right)
    \left(
    \begin{array}{c}
         \frac{1}{2} e^{-i \vartheta /2}
        \Big(
        \gamma_{2i - 1, \uparrow}
        -
        i
        \gamma_{2i, \uparrow}
        \Big)
        \\
        -\frac{i}{2} e^{-i \vartheta /2}
        \Big(
        \gamma_{2i - 1, \downarrow}
        -
        i
        \gamma_{2i, \downarrow}
        \Big)
    \end{array}
    \right)
    \\
    &=
    \left(
    \begin{array}{c}
        \frac{1}{2}\cos
        \left(
        - \frac{\pi}{4} + \frac{\tilde{\theta}(x_i)}{2}
        \right)
        e^{-i (\vartheta + \tilde{\phi}(x_i))/2}
        \left(\gamma_{2i-1, \uparrow} - i \gamma_{2i, \uparrow} \right)
        -
        \frac{i}{2} 
        \sin
        \left(
        - \frac{\pi}{4} + \frac{\tilde{\theta}(x_i)}{2}
        \right)
        e^{-i (\vartheta - \tilde{\phi}(x_i))/2}
        \left(\gamma_{2i-1, \uparrow} - i \gamma_{2i, \uparrow} \right)
        \\
        \frac{1}{2}\sin
        \left(
        - \frac{\pi}{4} + \frac{\tilde{\theta}(x_i)}{2}
        \right)
        e^{-i (\vartheta + \tilde{\phi}(x_i))/2}
        \left(\gamma_{2i-1, \downarrow} - i \gamma_{2i, \downarrow} \right)
        +
        \frac{i}{2} 
        \cos
        \left(
        - \frac{\pi}{4} + \frac{\tilde{\theta}(x_i)}{2}
        \right)
        e^{-i (\vartheta - \tilde{\phi}(x_i))/2}
        \left(\gamma_{2i-1, \downarrow} - i \gamma_{2i, \downarrow} \right)
    \end{array}
    \right).
    \label{appendixEq:c_in_terms_of_gamma.general_d}
\end{align}
Here, the local $d$ vector orientation is encoded in $U(x_i)$.
Substituting Eq.~\eqref{appendixEq:c_in_terms_of_gamma.general_d} into Eq.~\eqref{appendixEq:kitaev_and_sd} yields the Hamiltonian for a spatially varying $d$ vector in basis of Majorana operators.

\begin{figure}
    \centering
    \includegraphics[width= \linewidth]{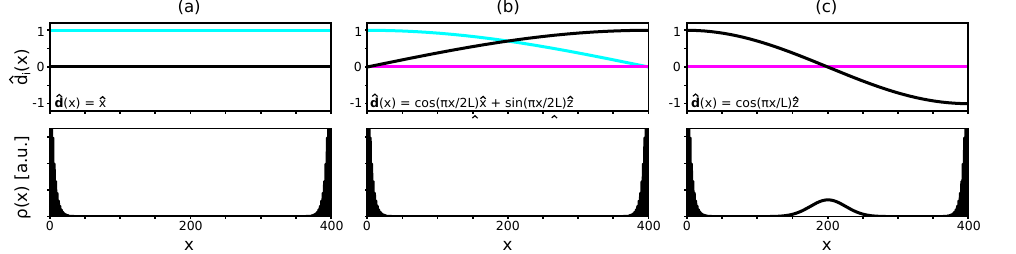}
    \caption{Zero-energy Majorana bound states for varying $d$ vector textures for the one-dimensional spinful Kitaev chain.
    The $x$, $y$, and $z$ components of the real unitary $d$ vector are shown in cyan, magenta, and black respectively in the top panels, and the real-space profile of the zero-energy states, $\rho(x_i)$, is shown in the bottom panels.
    \textbf{(a)}~Constant $d$ vector $\hat{\mathbf{d}}(x_i) = \hat{\mathbf{x}}$ gives rise to four degenerate zero-energy states, with two localized zero-energy states at each edge of the chain.
    \textbf{(b)}~A smoothly twisting $d$ vector with constant magnitude, $\hat{\mathbf{d}}(x_i) = \cos( \pi x_i/ 2L) \hat{\mathbf{x}} + \sin( \pi x_i/ 2L) \hat{\mathbf{z}}$ , similarly gives rise to four degenerate zero-energy states. 
    \textbf{(c)}~A $d$ vector texture containing a single domain wall in which the gap closes and changes sign, $\hat{\mathbf{d}}(x_i) = \cos(\pi x_i/L) \hat{\mathbf{z}},$ yields eight zero-energy states.
    In addition to the four states localized at both edges of the chain, there are four additional states localized at the domain wall at $x = L/2$.
    System parameters are $\mu = 0$ and $\Delta_0 = |t_0|/10$, and the chain has $400$ sites.
    }
    \label{fig:majorana_states}
\end{figure}

For spatially varying $d$ vector texture, it follows that the edge states are localized at the edge of the sample, provided that the bulk pairing gap does not close.
As an example, we consider a spatially varying $d$ vector in which
the $p_x$-wave orbital symmetry of the pairing order is retained.
From the bulk-edge correspondence, the $p_x$-wave pairing symmetry leads to topologically protected edge-state, while the $d$ vector texture determines their spin structure.
We consider two possible cases, in which the bulk pairing gap remains finite or if it closes and changes sign.

In the first case, if the superconducting gap does not close, the zero-energy edge states are still localized at the edge of the sample, as shown in Fig.~\ref{fig:majorana_states}(b).
As an example, we consider $d$ vector texture given by $\hat{\mathbf{d}}(x_i) = \cos(\pi x_i/2L)\hat{\mathbf{x}} + \sin(\pi x_i/2L)\hat{\mathbf{z}}$, which smoothly twists from pointing along the $x$ to the $z$ direction.
There are in total four zero-energy bound states, with the only difference from the standard Kitaev chain being the pairing spin structure, which is determined by the local $d$ vector texture.
The bulk topological invariant is determined by the orbital pairing symmetry and is independent of the orientation of the $d$ vector, provided that the gap does not close.
Consequently, the bulk-edge correspondence guarantees a fixed number of Majorana zero modes at the boundary, even in the presence of a spatially varying $d$ vector texture.

In the second case, we consider when the bulk gap closes and changes sign.
As a representative example, consider the pairing gap function described by the $d$ vector
\begin{math}
    \hat{\mathbf{d}}(x_i) = \cos(\pi x_i/L) \hat{\mathbf{z}},
\end{math}
in which $L$ is the length of the chain.
In this case, there are eight total zero-energy states.
In addition to the four edge states localized to the edge of the junction, there are four states that are localized at $x= L/2$, where the pairing gap vanishes and changes sign, as shown in Fig.~\ref{fig:majorana_states}(c).
This represents a topological domain wall, analogous to the Jackiw-Rebbi mode~\cite{Jackiw1976}.
By extension, for pairing gap function described by $\hat{\mathbf{d}}(x_i) = \cos(2\pi x_i/N)$, which closes and reopens the gap twice, there are twelve zero-energy gap states, with the additional four coming from states localized at the additional domain wall.


\subsection{Contribution to Josephson tunneling from Majorana states}

Lastly, we outline the contribution from Majorana states to Josephson tunneling and diode effect.
Consider an SIS junction between two one-dimensional $p_x$ wave superconductors of $N$ sites, with pairing orders described by $d$ vectors $\hat{\mathbf{d}}_L$ and $\hat{\mathbf{d}}_R$ respectively.
The tunneling across the interface between the left and right chains is described by
\begin{equation}
    H_{T} = 
    \sum_{\alpha, \alpha'} \left(  c^\dagger_{L, N, \alpha} T_{\alpha, \alpha'} c_{R, 1, \alpha}
    + \mathrm{h.c.} \right),
\end{equation}
in which $T_{\alpha \alpha'}$ is the tunneling matrix.
Without loss of generality, suppose that $\hat{\mathbf{d}}_L \parallel \hat{\mathbf{x}}$, and $\hat{\mathbf{d}}_R$ is related by the transformation in Eq.~\eqref{appendixEq:rotation}.
Consider projecting the tunneling to the unpaired Majorana states at the junction interface.
From Eq.~\eqref{appendixEq:c_in_terms_of_gamma.d_along_x}, the projected operators for the left side are given by
\begin{equation}
    P
    \left(
    \begin{array}{c}
         \tilde{c}_{N, \uparrow}^\dagger
         \\
         \tilde{c}_{N, \downarrow}^\dagger
    \end{array}
    \right)
    =
    \frac{1}{2}
    \left(
    \begin{array}{c}
         -i e^{-i \vartheta_L/2} \gamma_{2N, \uparrow}
         \\
         -e^{-i \vartheta_L/2} \gamma_{2N, \downarrow}
    \end{array}
    \right),
\end{equation}
and from Eq.~\eqref{appendixEq:c_in_terms_of_gamma.general_d}, the projected operators for the right side are given by
\begin{equation}
    P
    \left(
    \begin{array}{c}
         \tilde{c}_{N+1, \uparrow}
         \\
         \tilde{c}_{N+1, \downarrow}
    \end{array}
    \right)
    =
        \left(
    \begin{array}{c}
        \frac{1}{2}\cos
        \left(
        - \frac{\pi}{4} + \frac{\tilde{\theta}}{2}
        \right)
        e^{i (\vartheta_R + \tilde{\phi})/2}
        \gamma_{2N+1, \uparrow}
        -
        \frac{i}{2} 
        \sin
        \left(
        - \frac{\pi}{4} + \frac{\tilde{\theta}}{2}
        \right)
        e^{i (\vartheta_R - \tilde{\phi})/2}
        \gamma_{2N+1, \downarrow}
        \\
        \frac{1}{2}\sin
        \left(
        - \frac{\pi}{4} + \frac{\tilde{\theta}}{2}
        \right)
        e^{i (\vartheta_R + \tilde{\phi})/2}
        \gamma_{2N+1, \uparrow}
        +
        \frac{i}{2} 
        \cos
        \left(
        - \frac{\pi}{4} + \frac{\tilde{\theta}}{2}
        \right)
        e^{i (\vartheta_R - \tilde{\phi})/2}
        \gamma_{2N+1, \downarrow}
    \end{array}
    \right).
\end{equation}
Here, $\vartheta_L$ and $\vartheta_R$ are the $\mathrm{U}(1)$ superconducting phases of the left and right side respectively.
The projected tunneling Hamiltonian is given by
\begin{equation}
    P H_T P = \sum_{\alpha, \alpha' = \uparrow, \downarrow}
    e^{i (\vartheta_R - \vartheta_L)/2}
    \gamma_{2N, \alpha} T^\mathrm{proj}_{\alpha, \alpha'} \gamma_{2N+1, \alpha'}
    +
    \mathrm{h.c.}
    \label{appendixEq:projected_tunneling_majorana}
\end{equation}
in which $T^\mathrm{proj}_{\alpha, \alpha'}$ is the projected tunneling matrix.
For the case of spin-independent tunneling, $T = t_0 \sigma_0$ with $t_0$ being the tunneling amplitude, the projected tunneling matrix is given by
\begin{equation}
    T^\mathrm{proj}
    =
    \frac{t_0}{2}
    \left(
    \begin{array}{cc}
        -i\cos
        \left(
        - \frac{\pi}{4} + \frac{\tilde{\theta}}{2}
        \right)
        e^{i \tilde{\phi}/2}
        &
        -
        \sin
        \left(
        - \frac{\pi}{4} + \frac{\tilde{\theta}}{2}
        \right)
        e^{-i \tilde{\phi}/2}
        \\
        -\sin
        \left(
        - \frac{\pi}{4} + \frac{\tilde{\theta}}{2}
        \right)
        e^{i \tilde{\phi}/2}
        &
        -i \cos
        \left(
        - \frac{\pi}{4} + \frac{\tilde{\theta}}{2}
        \right)
        e^{-i \tilde{\phi}/2}
    \end{array}
    \right).
\end{equation}
The $4\pi$-periodic Josephson current can be seen from the $\mathrm{U}(1)$ phase difference in Eq.~\eqref{appendixEq:projected_tunneling_majorana}.

The Josephson supercurrent can generally be expressed as a superposition of different order contributions,
\begin{equation}
    I_\mathrm{tot}(\vartheta_{RL}) = 
    I_{1/2} \sin \Big( \frac{1}{2}(\vartheta_{RL} - \varphi_{1/2}) \Big)
    +
    \sum_{n=1,2, 3, \cdots} I_{n} \sin \Big( n (\vartheta_{RL} - \varphi_n) \Big)
\end{equation}
Here, $\vartheta_{RL} = \vartheta_R - \vartheta_L$ is the $\mathrm{U}(1)$ phase difference of the pairing orders at the left and right sides of the junction, and  $I_n$ and $\varphi_n$ are the magnitude and phase offset of the critical current for the $n\mathrm{th}$ order contribution, with $n = 1/2$ denoting $4\pi$-periodic Josephson current from the Majorana states localized at the junction interface.
Following the standard derivation of the Josephson diode effect~\cite{Wu2022, Davydova2022, Zhang2022a, Wang2025, Frazier2025b}, the superposition of different order Josephson currents with different phase offsets can lead to a nonvanishing diode effect.
Consequently, the $4\pi$-periodic Josephson current arising from the Majorana states can enhance or suppress the diode effect, depending on the system parameters and effective tunneling at the junction interface.
Nonetheless, the underlying features of the diode effect, which originate from breaking time-reversal and parity symmetries, persist.

}

%% file: supplemental/s8.large_Jsd.tex
\section{Effective tunneling in the limit of strong \texorpdfstring{$s$-$d$}{s-d} exchange}

\label{appendix:strong_Jsd_tunneling}

{\color{black}

In this section, we describe the effective tunneling in the presence of $s$-$d$ exchange in the $J_{sd} \gg t_0$ limit.
Consider the $s$-$d$ model
\begin{align}
    H 
    &=
    -\mu
    \sum_{i, \alpha} c^\dagger_{i, \alpha} c_{i, \alpha}
    +
    t_0 \sum_{\langle ij\rangle, \alpha} c^\dagger_{i, \alpha} c_{j, \alpha}
    +
    J_{sd} \sum_{i, \alpha, \beta}
    c^\dagger_{i, \alpha} \left[ \mathbf{s}_i \cdot \boldsymbol{\sigma} \right]_{\alpha \beta} c_{i, \beta},
\end{align}
as detailed in Sec.~\ref{SM:sd_models}.
In the limit of strong exchange coupling $J_{sd} \gg t_0$, the electrons are polarized according to the local exchange field $\hat{\mathbf{s}}_i$.
The state at the $i\mathrm{th}$ site is represented by
\begin{equation}
    | \hat{\mathbf{n}}_i\rangle
    =
    \left(
    \begin{array}{c}
         \cos \frac{\theta_i}{2}
         \\
         e^{i \varphi_i} \sin \frac{\theta_i}{2}
    \end{array}
    \right)
    \label{strong_J.eigenstates}
\end{equation}
up to a $\mathrm{U}(1)$ phase.
Here, $\hat{\mathbf{n}}_i = \langle \hat{\mathbf{n}}_i | \boldsymbol{\sigma} | \hat{\mathbf{n}}_i \rangle = 
( \sin \theta_i \cos \varphi_i, \sin \theta_i \sin \varphi_i, \cos \theta_i )^\mathrm{T}$
is the local magnetization of the itinerant electron at site $i$.
For antiferromagnetic coupling ($J_{sd} > 0$), $\hat{\mathbf{n}}_i = -\hat{\mathbf{s}}_i$, whereas for ferromagnetic coupling ($J_{sd} < 0$), $\hat{\mathbf{n}}_i = \hat{\mathbf{s}}_i$.
Treating the spin-independent nearest neighbour hopping perturbatively, the effective hopping matrix element between neighboring sites $i$ and $j$ as~\cite{Ye1999, Ohgushi2000}
\begin{equation}
    t_{ij}^\mathrm{eff} = t_0 \langle \hat{\mathbf{n}}_i | \hat{\mathbf{n}}_j\rangle
    =
    t_0 |\langle \hat{\mathbf{n}}_i | \hat{\mathbf{n}}_j\rangle|
    e^{i \mathrm{arg}(\langle \hat{\mathbf{n}}_i | \hat{\mathbf{n}}_j\rangle)}
    \label{appendixEq:effective_t_ij.band_basis}
\end{equation}
in which
\begin{equation}
    |\langle \hat{\mathbf{n}}_i | \hat{\mathbf{n}}_j\rangle|
    =
    \sqrt{\frac{1 + \hat{\mathbf{n}}_i \cdot \hat{\mathbf{n}}_j}{2}}.
\end{equation}
The complex phase arises from the geometric gauge field, with
\begin{math}
    \arg (\langle \hat{\mathbf{n}}_i | \hat{\mathbf{n}}_j\rangle)
    \sim
    \int_i^j \mathbf{a}(\boldsymbol{\ell}) \cdot \mathrm{d}\boldsymbol{\ell},
\end{math}
with $\mathbf{a}(\boldsymbol{\ell})$ being the vector potential, and can lead to an anomalous quantum Hall effect for noncoplanar spin configurations~\cite{Nagaosa2010}.

Next, we write the effective tunneling in the $J_{sd} \gg t_0$ limit in the spin-basis.
In the band-diagonal basis, the hopping matrix element is given by
\begin{equation}
    T_{ij}^{(b)} 
    =
    t_0
    \left(
    \begin{array}{cc}
         \langle \hat{\mathbf{n}}_i|\hat{\mathbf{n}}_j\rangle
         &
         \langle \hat{\mathbf{n}}_i|-\hat{\mathbf{n}}_j\rangle
         \\
         \langle -\hat{\mathbf{n}}_i|\hat{\mathbf{n}}_j\rangle
         &
         \langle -\hat{\mathbf{n}}_i|-\hat{\mathbf{n}}_j\rangle
    \end{array}
    \right).
\end{equation}
The projected hopping matrix between nearest neighbors $i$ and $j$ in the band-diagonal basis is given by
\begin{align}
    PT_{ij}^{(b)}
    =
    \left(
    \begin{array}{cc}
         t^\mathrm{eff}_{ij}
         & 0  
         \\
         0 & 0
    \end{array}
    \right),
\end{align}
with $t_{ij}^\mathrm{eff}$ given in Eq.~\eqref{appendixEq:effective_t_ij.band_basis}.
As such, the projected hopping matrix, in the spin-basis, is given by
\begin{equation}
    T_{ij}^\mathrm{eff}
    =
    t_0 \langle \hat{\mathbf{n}}_i| \hat{\mathbf{n}}_j\rangle
    \left(
    \begin{array}{cc}
        \langle 
        \uparrow \!
        |
        \hat{\mathbf{n}}_i
        \rangle
        \langle \hat{\mathbf{n}}_j
        | \!
        \uparrow
        \rangle
        &
        \langle 
        \uparrow \!
        |
        \hat{\mathbf{n}}_i
        \rangle
        \langle \hat{\mathbf{n}}_j
        | \!
        \downarrow
        \rangle
        \\
        \langle 
        \downarrow \!
        |
        \hat{\mathbf{n}}_i
        \rangle
        \langle \hat{\mathbf{n}}_j
        | \!
        \uparrow
        \rangle
        &
        \langle 
        \downarrow \!
        |
        \hat{\mathbf{n}}_i
        \rangle
        \langle \hat{\mathbf{n}}_j
        | \!
        \downarrow
        \rangle
    \end{array}
    \right).
\end{equation}
This can be decomposed into Pauli matrices $T_{ij}^\mathrm{eff} = T_{ij; 0}^\mathrm{eff} + \mathbf{T}_{ij}^\mathrm{eff}\cdot\boldsymbol{\sigma}$, in which
\begin{equation}
    \begin{aligned}
        T_{ij; 0}^\mathrm{eff}
        &= 
        \frac{t_0}{2}
        \langle \hat{\mathbf{n}}_i| \hat{\mathbf{n}}_j\rangle
        \Big(
        \langle \hat{\mathbf{n}}_j
        | \!
        \uparrow
        \rangle
        \langle 
        \uparrow \!
        |
        \hat{\mathbf{n}}_i
        \rangle
        +
        \langle \hat{\mathbf{n}}_j
        | \!
        \downarrow
        \rangle
        \langle 
        \downarrow \!
        |
        \hat{\mathbf{n}}_i
        \rangle
        \Big)
        \\
        T_{ij;x}^\mathrm{eff}
        &=
        \frac{t_0}{2}
        \langle \hat{\mathbf{n}}_i| \hat{\mathbf{n}}_j\rangle
        \Big(
        \langle \hat{\mathbf{n}}_j
        | \!
        \downarrow
        \rangle
        \langle 
        \uparrow \!
        |
        \hat{\mathbf{n}}_i
        \rangle
        +
        \langle \hat{\mathbf{n}}_j
        | \!
        \uparrow
        \rangle
        \langle 
        \downarrow \!
        |
        \hat{\mathbf{n}}_i
        \rangle
        \Big)
        \\
        T_{ij;y}^\mathrm{eff}
        &=
        \frac{it_0}{2}
        \langle \hat{\mathbf{n}}_i| \hat{\mathbf{n}}_j\rangle
        \Big(
        \langle \hat{\mathbf{n}}_j
        | \!
        \downarrow
        \rangle
        \langle 
        \uparrow \!
        |
        \hat{\mathbf{n}}_i
        \rangle
        -
        \langle \hat{\mathbf{n}}_j
        | \!
        \uparrow
        \rangle
        \langle 
        \downarrow \!
        |
        \hat{\mathbf{n}}_i
        \rangle
        \Big)
        \\
        T_{ij;z}^\mathrm{eff}
        &=
        \frac{t_0}{2}
        \langle \hat{\mathbf{n}}_i| \hat{\mathbf{n}}_j\rangle
        \Big(
        \langle \hat{\mathbf{n}}_j
         | \!
         \uparrow
         \rangle
         \langle 
         \uparrow \!
         |
         \hat{\mathbf{n}}_i
         \rangle
         -
         \langle \hat{\mathbf{n}}_j
         | \!
         \downarrow
         \rangle
         \langle 
         \downarrow \!
         |
         \hat{\mathbf{n}}_i
         \rangle
        \Big).
    \end{aligned}
    \label{appendixEq:effective_tunneling.stong_Jsd_limit.decomposition}
\end{equation}
Considering the summation of tunneling processes at the boundary between two superconducting grains, the effective tunneling matrices in Eq.~\eqref{appendixEq:effective_tunneling.stong_Jsd_limit.decomposition} correspond to the tunneling matrices in Eq.~\eqref{appendixEq:form_factor}.

Notably, a frustrated magnetic texture is not necessary to achieve the anisotropic Josephson couplings in Eq.~\eqref{josephson_coupling.d_vectors} in the limit of strong $s$-$d$ coupling.
For example, for a collinear ferromagnetic spin configuration, in which local spins at sites $i$ and $j$ are given by $\hat{\mathbf{s}}_i = \hat{\mathbf{s}}_j = -\mathrm{sgn}(J_{sd}) \hat{\mathbf{n}}_0$, the effective tunneling reduces to
\begin{equation}
    T^\mathrm{eff}_{ij} = \frac{t_0}{2}
    ( \sigma_0 + \hat{\mathbf{n}}_0\cdot \boldsymbol{\sigma}).
    \label{appendixEq:effective_tunneling.strong_Jsd.parallel_spins}
\end{equation}
The effective tunneling satisfies the condition in Eq.~\eqref{appendixEq:dm_coupling_condition}, which can give rise to a Dzyaloshinskii-Moriya-like Josephson coupling.
In contrast to the cases discussed in the main text, which consider $J_{sd} \ll t_0$, strong $s$-$d$ exchange can give rise to comparable amplitudes of the Josephson couplings $J_{nm}$, $\mathbf{D}_{nm}$, and $\Gamma_{nm}$ and can further promote a spatially inhomogeneous $d$ vector texture.
This regime can be pertinent to systems such as proximitized Mn$_3$Ge, in which the $s$-$d$ coupling can be comparable to the hopping amplitude~\cite{Chen2014, Kimata2019}.

As an illustrative example, consider Josephson tunneling in the presence of a ferromagnetic spin texture, in which the tunneling is given by that in Eq.~\eqref{appendixEq:effective_tunneling.strong_Jsd.parallel_spins}.
From Eq.~\eqref{appendixEq:josephson_couplings},
the Josephson couplings are given by
\begin{subequations}
    \begin{align}
        J_{nm} 
        &\approx
        - \Delta_0 \frac{t_0^2}{2W^2}
        N_{nm}
        \\
        \mathbf{D}_{nm}
        & \approx
        i\Delta_0 \frac{t_0^2}{2W^2}
        N_{nm}
        \hat{\mathbf{n}}_0
        \\
        \Gamma_{nm}^{ab}
        &\approx
        \Delta_0  \frac{t_0^2}{2 W^2} N_{nm} \hat{n}_0^a \hat{n}_0^b
    \end{align}
\end{subequations}
in which $N_{nm}$ is the dimensionless length of the interface between superconducting grains $n$ and $m$.
Here, the Heisenberg-like and DM-like Josephson couplings are of similar magnitude.
The diode efficiency for unitary pairing orders is given by~\cite{Frazier2025b}
\begin{equation}
    \eta \approx
    \frac{N_{nm}\Delta_0t_0^4 (\hat{\mathbf{d}}_n \cdot \hat{\mathbf{d}}_m)^2}{W^4 |\tilde{\mathcal{J}}_{nm}|} |\sin (2\phi_1)|
\end{equation}
in which we have taken $I_2 \approx (e/\hbar) \Delta_0 t_0^4/W^4$ and $\phi_2 \approx 0$ for simplicity, and $\tilde{\mathcal{J}}_{nm}$ is defined as
\begin{equation}
    \tilde{\mathcal{J}}_{nm} = 
    J_{nm}
    \hat{\mathbf{d}}_n \cdot \hat{\mathbf{d}}_m
    +
    \mathbf{D}_{nm}
    \cdot
    (
    \hat{\mathbf{d}}_n \times \hat{\mathbf{d}}_m)
    +
    \hat{\mathbf{d}}_n \Gamma_{nm} \hat{\mathbf{d}}_m,
    \label{appendixEq:tilde_J.phase_independent}
\end{equation}
Here, the phase difference in the first order critical current is given by~\cite{Frazier2025b}
\begin{equation}
    \tan \phi_1 \approx
    \frac{\hat{\mathbf{n}}_0 \cdot (\hat{\mathbf{d}}_n \times \hat{\mathbf{d}}_m)}{\hat{\mathbf{d}}_n \cdot \hat{\mathbf{d}}_m - (\hat{\mathbf{n}}_0\cdot\hat{\mathbf{d}}_n)
    (\hat{\mathbf{n}}_0\cdot\hat{\mathbf{d}}_m)}.
    \label{appendixEq:diode_efficiency_FM_interface}
\end{equation}
For collinear $d$ vectors, this yields $\phi_1 = 0$ and thus a vanishing diode effect.
However, for noncollinear $d$ vectors, this leads to finite phase offset $\phi_1$, resulting in a finite diode effect, with the diode efficiency largely being determined by the relative alignment of $d$ vectors.
In contrast to the diode effect in Eq.~\eqref{nonreciprocal_current}, which is primarily determined by the underlying exchange field, the diode effect in Eq.~\eqref{appendixEq:diode_efficiency_FM_interface} is primarily dependent on the relative orientation of $d$ vectors at the two sides of the junction.
For example, consider the case that $\hat{\mathbf{d}}_n \cdot \hat{\mathbf{d}}_m = \cos \theta$, and both $\hat{\mathbf{d}}_n$ and $\hat{\mathbf{d}}_m$ are orthogonal to $\hat{\mathbf{n}}_0$ such that $\hat{\mathbf{n}}_0\cdot (\hat{\mathbf{d}}_n \times \hat{\mathbf{d}}_m) = \sin \theta.$
It follows that
\begin{equation}
    \eta_\mathrm{max} \approx
    \frac{2 t_0^2}{W^2} \cos^2 \theta |\sin 2\theta|,
\end{equation}
which has a maximum of $\eta \approx \frac{3 \sqrt{3}}{4} \frac{t_0^2}{W^4}$ at $\theta = \pm \frac{\pi}{6}, \pm \frac{7 \pi}{6}$.
}

%% file: supplemental.bbl
%